\documentclass[11pt]{article}%\pdfoutput=1
\usepackage{cite}
\usepackage{scalerel}
\usepackage{shuffle}
\usepackage{amsmath,amsfonts,amssymb}
\usepackage{amsmath,amssymb,amsthm}
\usepackage{latexsym,epsfig}
\usepackage{dsfont}
\usepackage{hyperref}
\usepackage{extarrows}
\usepackage{comment} % To comment several lines at once.
\usepackage{tikz-cd}

\usepackage{tikz}
\usetikzlibrary{arrows.meta}

\usepackage{mathtools}

\usepackage{listings}
\usepackage{xcolor}
\usepackage[T1]{fontenc}

\definecolor{leanKW}{RGB}{125, 30, 130}   % keywords (purple)
\definecolor{leanCO}{RGB}{0, 100, 0}       % comments (green)
\definecolor{leanBG}{RGB}{248, 248, 248}   % background (light gray)
\definecolor{leanFR}{RGB}{200, 200, 200}   % frame

\newtheoremstyle{colon}{\topsep}{\topsep}{\itshape}{}{\bfseries}{:}{.5em}{}
\theoremstyle{colon}

\newtheorem{prop}{Proposition}

\newcommand{\dbar}{\bar\partial}

\newcommand{\PV}{\mathrm{PV}}

\newcommand{\Lie}{\mathcal L}
\newcommand{\vth}{\vartheta}
\newcommand{\dv}{\operatorname{div}_\omega}
\newcommand{\cip}[2]{\langle\!\langle #1, #2\rangle\!\rangle}
\newcommand{\Kk}{\mathcal K}

\def\hybrid{
        \topmargin -20pt
        \oddsidemargin 0pt
        \headheight 0pt \headsep 0pt
        \textwidth 6.25in % A4 paper
        \textheight 9.5in % A4 paper
        \marginparwidth .875in
        \parskip 5pt plus 1pt \jot = 1.5ex}

\hybrid

 \csname
@addtoreset\endcsname{equation}{section}

\def\be{\begin{equation}}
\def\ee{\end{equation}}

\allowdisplaybreaks

\def\bpm{\begin{pmatrix}}
\def\epm{\end{pmatrix}}

\renewenvironment{proof}[1][Proof]{\par\noindent\textit{#1:}\ }{\hfill$\square$\par\medskip}
\begin{document}

\begin{titlepage}
\rightline{}
%\rightline\today
\rightline{October 2026}
%\rightline{HU-EP-25/39-RTG}  
\begin{center}
\vskip 1.5cm
{\Large \bf{A local action for Kodaira--Spencer gravity
 }}
\vskip 1.2cm

{\large\bf {Olaf Hohm$^1$ }} \vskip 1.1cm

$^1$ {\it  Institute for Physics, Humboldt University Berlin,\\
 Zum Gro\ss en Windkanal 2, D-12489 Berlin, Germany}\\[2.5ex]

ohohm@physik.hu-berlin.de 

\vskip .1cm

\vskip .2cm

\end{center}

\bigskip\bigskip
\begin{center} 
\textbf{Abstract}

\end{center} 
\begin{quote}

A local action is given for Kodaira--Spencer gravity by solving  a constraint on the 
complex structure deformation of a Calabi--Yau threefold  in terms of a $(1,1)$ form as the new gravity field.  
The gauge algebra becomes a genuine $L_{\infty}$ algebra based on a 
Courant algebroid, while the gravity field and its linearized gauge transformations coincide with those of double field theory. 
On a real three-dimensional slice, the free action reduces to the Fierz--Pauli action for massless spin-2 plus the Kalb--Ramond 
action for a B-field. Non-linearly, the theory carries parity-odd interactions with three derivatives, and 
it is gauge invariant without a level-matching constraint, 
in this sense defining  an unconstrained double field theory. 
The full theory is encoded in a cyclic $L_{\infty}$ algebra, 
for which a double copy interpretation is given.

\end{quote} 
\vfill
\setcounter{footnote}{0}
\end{titlepage}

\tableofcontents
%\newpage

\section{Introduction}

Based on the deformation theory of complex structures going back to Kodaira and Spencer \cite{KS58}, 
Kodaira--Spencer gravity is a topological field theory introduced by Bershadsky, Cecotti, Ooguri and Vafa (BCOV) 
as the closed string field theory of the so-called topological B-model \cite{BCOV93,BCOV}, 
whose open string counterpart is the holomorphic Chern--Simons theory of \cite{Witten92}.
The  perturbative quantization of BCOV, which has been formulated in the Batalin--Vilkovisky (BV) formalism  by
Costello and Li \cite{CL,CL2}, yields the higher genus amplitudes of topological string theory. 
(See also \cite{CL3} for a proposal to obtain Kodaira--Spencer gravity from a twist of
type IIB supergravity, which in turn features in `twisted holography' \cite{CG}.)

The goal of this paper is to present a local action principle for BCOV theory or Kodaira--Spencer gravity. 
This theory is defined on a Calabi--Yau threefold $X$, and the gravity field is given by the complex structure 
deformation (also known as Beltrami differential), which in local coordinates is given by the mixed tensor 
$A=A_{\bar\imath}{}^{j}(z,\bar{z})\, d\bar{z}^{\bar\imath}\otimes \frac{\partial}{\partial z^j}$, 
where $z$ and $\bar{z}$ are holomorphic and antiholomorphic coordinates, respectively.  
The linearized field equations state that $A$ is closed with respect to the antiholomorphic Dolbeault differential, 
 \begin{equation}\label{origKSeq}
  \bar{\partial}A=0\;. 
 \end{equation} 
These equations cannot be obtained from a local action principle for $A$, but one can write a non-local action upon 
imposing a constraint as follows.  Any Calabi--Yau threefold carries a holomorphic $(3,0)$ form with 
components $\omega_{ijk}$, which can be used to define the $(2,1)$ form 
$A'_{ij\,\bar k} = \omega_{ijl} \,A_{\bar k}{}^{l}$. In terms of this we impose the constraint $\partial A'=0$, 
which is equivalent to  the  original Beltrami differential being  divergence-free in its 
holomorphic vector index, i.e., $\partial_{j}A_{\bar\imath}{}^{j}=0$. One can then show that $\bar{\partial}A'$ is $\partial$ 
exact, i.e., $\bar{\partial}A'=\partial\Psi$ for a $(1,2)$ form $\Psi$. This defines the non-local operator 
$\partial^{-1}\bar{\partial}A':=\Psi$,  
in terms of which one can write the action 
 \begin{equation}
   S_{\mathrm{kin}}=\tfrac12\!\int_X A'\wedge\partial^{-1}\dbar A' \;. \label{eq:SkinNL} 
 \end{equation}
Variation with respect to $A'$ yields 
$\dbar A'=0$, which, owing 
to $\dbar$ commuting with the holomorphic $\omega$, in turn implies  the original Kodaira--Spencer gravity  equations \eqref{origKSeq}.

In this paper we use that the above action can be made local by solving the constraint $\partial A'=0$ in terms of a new $(1,1)$ form field 
$e=e_{i\bar\jmath}(z,\bar{z}) dz^i \wedge  d\bar{z}^{\bar\jmath}$:  
 \begin{equation}\label{eq:prepotINTRO}
   A'=\partial e 
   \qquad \Leftrightarrow \qquad A_{\bar\jmath}{}^{k} = \omega^{kli}\partial_l e_{i\bar\jmath}\;, 
 \end{equation}
as already observed by BCOV (see footnote 23 in  \cite{BCOV}). 
In order to see that this makes the quadratic term local we note 
with \eqref{eq:prepotINTRO} that $\dbar A'=-\partial\dbar e$, so  we can write $\partial^{-1}\dbar A'=-\dbar e$, 
and the kinetic term \eqref{eq:SkinNL} localizes to the two-derivative action 
\begin{equation}\label{eq:Skin}
S_{\mathrm{kin}} 
=-\tfrac12\!\int_X\partial e\wedge\dbar e \;. 
\end{equation}
Furthermore, Kodaira--Spencer gravity has  a cubic interaction term in terms of $A$ or, equivalently, $A'$  that is local and in fact 
carries no derivatives. 
We can hence write a local interaction term in terms of $e$ by simply replacing $A'$ according to \eqref{eq:prepotINTRO}, which yields
 cubic interaction terms with three derivatives.

 It is thus guaranteed that the entire action is local, but the interesting question is how 
 the gauge symmetries including diffeomorphisms are realized in this formulation.  Indeed, writing \eqref{eq:prepotINTRO} introduces a 
 gauge redundancy since  from this relation $e$ is only well-defined up to shifts $e\rightarrow e+\partial\bar{\lambda} $, with 
 $\bar{\lambda}=\bar{\lambda}_{\bar\imath}(z,\bar{z}) \,d\bar{z}^{\bar\imath}$ a $(0,1)$ form. Furthermore, standard Kodaira--Spencer gravity 
 is invariant under 
 volume-preserving holomorphic diffeomorphisms, with a parameter that can be encoded in 
 a $(1,0)$ form $\lambda=\lambda_{i}(z,\bar{z}) dz^i$, both
 acting on $e$ as\footnote{A closely related description appears in the work of Saberi and Williams \cite{Saberi:2020pmw}. Their $(1,1)$-form has the same linearized gauge transformations \eqref{linearizedGaugeINTRO} and is related to the Kodaira--Spencer field by $A'=\partial e$, but they do not 
 construct a local action or extend it non-linearly.} 
  \be\label{linearizedGaugeINTRO} 
   \delta e = \partial\bar{\lambda}  + \bar{\partial}\lambda \qquad \Leftrightarrow \qquad 
   \delta e_{i\bar\jmath} =  \partial_i\bar{\lambda}_{\bar\jmath}  - \bar{\partial}_{\bar\jmath}\lambda_i \;, 
  \ee
 where $\lambda$ and $\bar{\lambda}$ are independent parameters (and not complex conjugate to each other). 
The standard diffeomorphism action is with respect to the vector parameter 
 \be\label{anchoremnbedding} 
 \vartheta(\lambda)^i := \omega^{ijk} \partial_j\lambda_k\;, 
 \ee
 which is divergence-free and hence generates volume-preserving diffeomorphisms. 
 The non-linear gauge algebra turns out to be an intriguing higher $L_{\infty}$ and Leibniz algebra structure  based on a 
 Courant algebroid with a non-trivial three-bracket. Remarkably, precisely the same Leibniz algebras have appeared in 
 gauged supergravity \cite{deWit:2002vt,deWit:2008ta,Lavau:2014iva} 
 and double and exceptional field theory \cite{Hull:2009mi,Hull:2009zb,Hohm:2010jy,Hohm:2010pp,Coimbra:2011ky,Berman:2012vc,Hohm:2013pua,Hohm:2013vpa},
  with the `anchor' or `embedding tensor' being given by analogues 
 of \eqref{anchoremnbedding} \cite{Hohm:2018ybo,Kotov:2018vcz,Hohm:2019wql,Bonezzi:2019ygf,Lavau:2019oja,Bonezzi:2019bek}, see \cite{HS} for a closely related precursor. Indeed, in our formulation the perturbative 
 gravity field $e_{i\bar\jmath}$ is identical to the generalized metric fluctuation of double field theory, with the same linearized 
 gauge transformations \cite{Hull:2009mi,Hull:2009zb,Hohm:2010jy,Hohm:2010pp}. 
 Moreover, reducing to the three-dimensional (3D) real slice $z=\bar{z}$ and decomposing $e$ into its symmetric and antisymmetric 
 parts, the action \eqref{eq:Skin} yields the Fierz--Pauli action 
 for massless spin-2 plus the Kalb--Ramond action for a B-field. At higher order we obtain parity-odd higher-derivative couplings. 
 As in double field theory the 
 gauge algebra is a Courant algebroid, except that here no level-matching or `section' constraint, which would halve the number of 
 dimensions, needs to be imposed. In this sense, Kodaira--Spencer gravity is  an  unconstrained double field theory.

 The rest of this paper is organized as follows. In section 2 we give a self-contained review of Kodaira--Spencer gravity, 
 introducing the Dolbeault bicomplex, the Kodaira--Spencer equations and the non-local action. 
 The local reformulation of the action principle sketched above is developed in section 3. We begin by introducing the 
 Leibniz algebra underlying the Roytenberg--Weinstein construction of the $L_{\infty}$ algebra of Courant algebroids \cite{RW}, 
 which will form part of the (higher) gauge algebra.  Then we define a cyclic  $L_{\infty}$ algebra that encodes, in the 
 sense of \cite{Zwiebach:1992ie,Hohm:2017pnh},   the full 
 local reformulation of Kodaira--Spencer gravity and which is defined on a certain self-dual or `folded' subspace of 
 the Dolbeault complex. This cyclic $L_{\infty}$ algebra defines the complete BV master action, 
 which due to the higher gauge algebra includes quartic couplings in the ghost sector. 
 In subsection~\ref{sec:3d} we reduce the theory to a real 3D  slice and relate it to the Fierz--Pauli theory 
 and to an exotic deformation thereof found by Boulanger and Gualtieri \cite{BG}.
 In section 4 we give a `double copy' interpretation of the $L_{\infty}$ algebra that is analogous, though different 
 in important respects, to the homotopy algebraic double copy for which there is a growing literature 
 \cite{Bonezzi:2022bse,BCDH22,BCDH23,BCH24,BJKMSW} (see in particular  \cite{BDH} for a related 
 double copy discussion  of Kodaira--Spencer gravity). Conclusions and an outlook are in section 5,
 where we also compare with the recent works \cite{EKKS26} and \cite{BBZ26}, which relate
 Kodaira--Spencer gravity and its real analogue to three-dimensional gravity.  
In the bulk of this paper we assume that every $\partial$-closed $(2,1)$-form on $X$ is $\partial$-exact, so that
 $\partial A'=0$ can indeed be solved by \eqref{eq:prepotINTRO}. This still captures several interesting cases,
 including $X=\mathbb{C}^3$, but in subsection~\ref{sec:cohom} we show
 that it is relatively straightforward to relax this assumption by including the moduli corresponding
 to the harmonic pieces in the cohomology of a compact Calabi--Yau threefold.

\section{Non-local action for Kodaira--Spencer gravity}\label{sec:review}

In this section we review standard Kodaira--Spencer gravity. To keep the paper self-contained, 
in the first subsection we recall the basics 
of complex three-manifolds and introduce the Dolbeault complex of differential forms. The equations of motion 
of Kodaira--Spencer gravity are defined in the second subsection, based on the complex of antiholomorphic forms 
taking values in the holomorphic polyvectors. The final subsection introduces  the non-local action.

\subsection{Generalities}

Let  $X$ be a Calabi--Yau threefold. The reader may think of $X=\mathbb{C}^3$, but the 
following results will be valid for any Calabi--Yau threefold. We can view $\mathbb{C}^3$ 
as the real six-dimensional manifold $\mathbb{R}^6$ with coordinates 
 \be
  X^M = (x^i, y^i)\;, \quad i=1,2,3\;. 
 \ee 
We can alternatively  introduce complex coordinates 
 \be\label{DEFzandbarz}  
   z^i=x^i+i\,y^i\;, \qquad  \bar z^{\bar\imath}=x^i-i\,y^i\;,
 \ee 
where again $i=1,2,3$, and an $i$ that does not appear as an index is a complex unit  with $i^2=-1$. 
These formulas motivate the definition of the  following differential operators 
\begin{equation}\label{eq:coords}
\partial_i:=\frac{\partial}{\partial z^i}
=\frac12\Big(\frac{\partial}{\partial x^i}-i \frac{\partial}{\partial y^i}\Big)\;,
\qquad
\bar{\partial}_{\bar\imath}:=\frac{\partial}{\partial\bar z^{\bar\imath}}
=\frac12\Big(\frac{\partial}{\partial x^i}+i \frac{\partial}{\partial y^i}\Big)\,, 
\end{equation}
which obey $\partial_i z^j=\delta_i{}^{j}=\bar{\partial}_{\bar\imath} \bar{z}^{\bar\jmath}$ and 
$\partial_i \bar{z}^j=0=\bar{\partial}_{\bar\imath} z^j$. 

Next we define the space $\Omega^{p,q}$ of $(p,q)$-forms (the Dolbeault complex). A  $(p,q)$-form can be written as 
\be 
 \alpha=\frac1{p!q!}\alpha_{i_1\cdots i_p\,\bar\jmath_1\cdots\bar\jmath_q}(z,\bar{z}) \, 
  dz^{i_1}\!\wedge\cdots\wedge dz^{i_p} \wedge d\bar{z}^{\bar\jmath_1}\wedge\cdots \wedge d\bar z^{\bar\jmath_q}\;, 
\ee
whose components are separately antisymmetric in the first group of $p$ indices and the last group 
of $q$ indices. The familiar calculus of differential forms of a real manifold generalizes as follows: 
The de Rham differential is replaced by two differentials $\partial$ and $\bar{\partial}$ that act on functions as 
$\partial f=\partial_i f\,dz^i$ and $\dbar f=\bar{\partial}_{\bar\imath}f \,d\bar z^{\bar\imath}$  
and on arbitrary $(p,q)$ forms as 
\begin{equation}\label{eq:dops}
\partial=dz^i\wedge\partial_i\,:\; \Omega^{p,q}\to\Omega^{p+1,q},\qquad
\dbar=d\bar z^{\bar\imath}\wedge\bar{\partial}_{\bar\imath}\,:\; \Omega^{p,q}\to\Omega^{p,q+1}\;. 
\end{equation} 
These operators  square to zero just as the familiar de Rham differential does. 
Moreover, it follows by a straightforward computation that $\partial$ and $\bar{\partial}$ 
anticommute, so that we have in total 
\begin{equation}\label{eq:bicx}
\partial^2=0\,,\qquad \dbar^2=0\,,\qquad \partial\dbar+\dbar\partial=0 \;. 
\end{equation}
These relations in turn imply  that there is a total differential 
\begin{equation} 
\qquad d=\partial+\dbar \;, 
\end{equation}
that squares to zero, $d^2=0$. The total space of $(p,q)$ forms denoted by $\Omega^{\bullet,\bullet}$ then 
forms a \emph{Dolbeault bicomplex}: a bigraded space 
with two anticommuting derivations. Moreover, 
$(\Omega^{\bullet,\bullet},\wedge,\partial,\dbar)$ carries a bigraded-commutative algebra structure given by the wedge product $\wedge$. 

We will frequently use the so-called $\partial\dbar$-lemma, which relates the cohomologies of $\partial$, $\dbar$ and $d$.
On a compact K\"ahler manifold, and hence on a compact Calabi--Yau threefold, it states that for a
form $\alpha$ that is both $\partial$-closed and $\dbar$-closed the following conditions are equivalent: $\alpha$ is
$d$-exact, $\alpha$ is $\partial$-exact, $\alpha$ is $\dbar$-exact, and $\alpha=\partial\dbar\beta$ for some $\beta$
\cite{DGMS,Huybrechts}. 
The local version holds without compactness: locally, every $d$-closed $(p,q)$-form with $p,q\ge1$ is of
the form $\partial\dbar\beta$.

In the following the holomorphic polyvector fields  will also play an important role. 
These are fields with antisymmetric upper holomorphic indices whose   space we denote by $\wedge^pT^{1,0}$. 
In order to relate forms to polyvectors we will assume that there is a nowhere vanishing 
holomorphic $(3,0)$-form $\omega$. In local coordinates we write 
 \be 
  \omega=\frac{1}{3!}\, \omega_{ijk}\, dz^i\wedge dz^j\wedge dz^k \ \in \ \Omega^{(3,0)} \;, 
 \ee 
where the $\omega_{ijk}$ are totally antisymmetric.    
By degree reasons this form obeys   
$\partial\omega=0$, and we will also assume $\dbar\omega=0$. 
We further  introduce  the dual trivector  
 \be 
   \omega^{-1} = \frac{1}{3!}\, \omega^{ijk}\, \partial_i \wedge \partial_j \wedge \partial_k  \ \in \ \wedge^3T^{1,0}\;, 
 \ee  
whose 
components $\omega^{ijk}$ are also totally antisymmetric. 
For an antiholomorphic form taking values in the holomorphic polyvectors we can write 
 \be 
  \beta = \frac{1}{p!q!} \,\beta_{\bar\imath_1\ldots \bar\imath_q}{}^{j_1\ldots j_p}(z,\bar{z})   \,d\bar{z}^{\bar\imath_1}\wedge \cdots  \wedge d\bar{z}^{\bar\imath_q} \ \otimes \ 
  \partial_{j_1}\wedge \cdots \wedge \partial_{j_p} 
  \in\Omega^{0,q}(\wedge^pT^{1,0})\;, 
 \ee
 i.e., in local coordinates such a tensor is a mixed tensor, with unbarred upper indices and barred lower indices.  
 Given the nowhere vanishing 
 holomorphic $(3,0)$-form $\omega$ we have an  isomorphism mapping  $\beta \in\Omega^{0,q}(\wedge^pT^{1,0})$ to  
   $\beta\lrcorner\omega\in\Omega^{3-p,q}$, which is 
 given by the contraction of the upper (polyvector) indices with $\omega$, 
 \be\label{eq:contraction}
  \beta\lrcorner\omega=\frac{1}{p!\,(3-p)!\,q!}\,
  \beta_{\bar\imath_1\cdots\bar\imath_q}{}^{j_1\cdots j_p}\, 
  \omega_{j_1\cdots j_p\,k_1\cdots k_{3-p}}\,
  dz^{k_1}\!\wedge\cdots\wedge dz^{k_{3-p}}\wedge
   d\bar z^{\bar\imath_1}\!\wedge\cdots\wedge d\bar z^{\bar\imath_q}\;. 
 \ee

\subsection{Kodaira--Spencer theory}

We are now ready to define Kodaira--Spencer theory. It is defined on the space of antiholomorphic forms taking 
values in the holomorphic polyvector  fields, 
 \be
  \PV=\bigoplus_{i,j}\PV^{i,j}\;, \qquad \PV^{i,j}=\Omega^{0,j}(\wedge^iT^{1,0})\;,  
 \ee
where the degree is given by $i+j$ \cite{CL}. This space carries the wedge product $\wedge$, under which it is graded commutative with respect
to the total degree $|\beta|=i+j$ of $\beta\in\PV^{i,j}$. Here both the antiholomorphic one-forms
$d\bar z^{\bar\imath}$ and the vector fields $\partial_j$ are treated as odd, so that the wedge product of
$\beta\in\PV^{p,q}$ and $\gamma\in\PV^{p',q'}$ is computed by moving the $p$ vector fields of $\beta$ past the
$q'$ one-forms of $\gamma$, which gives a sign $(-1)^{pq'}$. It further carries the Dolbeault differential
$\dbar=d\bar z^{\bar\imath}\wedge\bar\partial_{\bar\imath}$, which acts from the left, and the \emph{divergence} operator
$\dv:\PV^{p,q}\to\PV^{p-1,q}$, which contracts the first upper index with a 
holomorphic derivative:\footnote{Strictly speaking, the most general expression of the holomorphic 3-form  is 
$\omega=f(z) dz^1\wedge dz^2 \wedge dz^3$, for which the divergence of a vector 
reads  ${\rm div}_{\omega}V=f^{-1}\partial_i(f V^i)$, and similarly for higher polyvectors, but for simplicity 
we assume $f=1$. The index-free definitions make it clear that all constructions go through  in general.}
\begin{equation}\label{eq:div}
(\dv\beta)_{\bar\imath_1\ldots\bar\imath_q}{}^{j_2\ldots j_p}=(-1)^q\,\partial_{j} \beta_{\bar\imath_1\ldots\bar\imath_q}{}^{jj_2\ldots j_p}\;,
\end{equation}
in terms of the components defined above. The sign $(-1)^q$ arises because $\dv$ acts from the left
and hence has to move past the $q$ antiholomorphic one-forms. Under the isomorphism with
forms the divergence corresponds to $\partial$ in the sense that 
\begin{equation}\label{eq:divomega}
(\dv\beta)\lrcorner\omega=(-1)^{|\beta|+1}\,\partial(\beta\lrcorner\omega)\;.
\end{equation}
$\dv$ squares to zero and is a \emph{second-order} operator for $\wedge$. This means that 
the triple
$(\PV,\wedge,\dv)$ is a BV algebra. This implies that the Schouten--Nijenhuis bracket
\begin{equation}\label{eq:koszul}
\{\beta,\gamma\}:=(-1)^{\beta}\big(\dv(\beta\wedge\gamma)
-\dv\beta\wedge\gamma-(-1)^{\beta}\beta\wedge\dv\gamma\big)\;, 
\end{equation}
where in exponents we write $\beta$ for $|\beta|$, is graded (anti-)symmetric, 
\be\label{eq:gradedsym}
 \{\beta,\gamma\}=-(-1)^{(\beta+1)(\gamma+1)}\{\gamma,\beta\}\;,
\ee
and satisfies the graded Jacobi identity {exactly}, 
\be\label{eq:Jacobi}
 \{\beta,\{\gamma,\alpha\}\}=\{\{\beta,\gamma\},\alpha\}+(-1)^{(\beta+1)(\gamma+1)}\{\gamma,\{\beta,\alpha\}\}\;. 
\ee
Moreover, $\bar{\partial}$ is a derivation of the bracket, 
 \be\label{LeibnizzzzzZZZ} 
\bar\partial\{\alpha,\beta\}=\{\bar\partial\alpha,\beta\}+(-1)^{|\alpha|+1}\{\alpha,\bar\partial\beta\}\;. 
 \ee 
These relations make in particular 
$(\PV,\wedge,\{\cdot,\cdot\})$ a Gerstenhaber algebra \cite{Getzler}.

In order to verify \eqref{LeibnizzzzzZZZ} explicitly one has to use that  $\dbar$ is a derivation of degree one of the wedge product,
$\dbar(\beta\wedge\gamma)=\dbar\beta\wedge\gamma+(-1)^{|\beta|}\beta\wedge\dbar\gamma$ and that  $\dbar$ and $\dv$
anticommute, 
 \begin{equation} 
   \dbar\,\dv+\dv\,\dbar=0\;. 
 \end{equation}   
 This relation in turn can be quickly verified in components or in index-free form  using  \eqref{eq:divomega}, for which one has to use 
 that for $\beta\in\PV^{p,q}$
\be\label{eq:dbaromega}
 (\dbar\beta)\lrcorner\omega=(-1)^{p+1}\,\dbar(\beta\lrcorner\omega)\;, 
\ee
since $\omega$ is holomorphic, so that $\dbar$ acts only on the components of $\beta$.

We are now ready to define Kodaira--Spencer gravity. 
To this end we note that  the subspace $({\rm PV}, \bar{\partial})$ forms a differential graded (dg) Lie algebra, 
which implies that we can postulate Maurer--Cartan equations as the field equations. These 
define the Kodaira--Spencer equations for the gravity field 
$A=A_{\bar\imath}{}^{j}(z,\bar{z})\, d\bar{z}^{\bar\imath}\otimes \frac{\partial}{\partial z^j}\in\PV^{1,1}$:
\begin{equation}\label{eq:KS}
\dbar A+\tfrac12\{A,A\}=0 \;.
\end{equation}
These equations are gauge covariant  under 
\be\label{eq:KSgauge}
\delta_\xi A=\bar\partial\xi+\{A,\xi\},\qquad
\xi\in\mathrm{PV}^{1,0}\;, 
\ee
with a vector gauge parameter $\xi^i$. 
Covariance  follows from $\bar{\partial}^2=0$, the Leibniz rule \eqref{LeibnizzzzzZZZ} and the Jacobi identity \eqref{eq:Jacobi}.

\subsection{Non-local action}\label{sec:nonlocal}

There is no local action for $A$ that implies the Kodaira--Spencer equations \eqref{eq:KS}, but one can define 
a non-local action under an assumption as follows: We  set 
 \be
   A'=A\lrcorner\omega\in\Omega^{2,1} \qquad \Leftrightarrow \qquad 
   A'_{ij\,\bar l} = \omega_{ijk} \,A_{\bar l}{}^{k}\;, 
 \ee
and then assume that this $(2,1)$ form is $\partial$ closed, 
 \be\label{A'Closed} 
  \partial A'= 0\;. 
 \ee
This is equivalent to the divergence $\dv A$ being zero (in local coordinates $\partial_jA_{\bar\imath}{}^{j}=0$). 
Indeed, by \eqref{eq:divomega} we have  $\partial A'=\partial(A\lrcorner\omega)=-(\dv A)\lrcorner\omega=-(\dv A)\,\omega$, where 
in the last step we used that $\dv A\in\PV^{0,1}$, and contraction of a scalar-valued
form with $\omega$ is multiplication. This constraint reduces the gauge transformations 
to `volume-preserving' diffeomorphisms with the vector parameter $\xi^i$ being divergence free, 
on which the dg Lie bracket closes.

We then use that for $\partial A'=0$ it follows  that  $\dbar A'$ is $\partial$-exact. 
Indeed, the $(2,2)$-form $\dbar A'$ is $\dbar$-closed and $\dbar$-exact, and it is $\partial$-closed since 
$\partial\dbar A'=-\dbar\partial A'=0$. By the $\partial\dbar$-lemma of section~\ref{sec:review} it is therefore of the
form $\dbar A'=\partial\dbar\gamma$ for some $\gamma\in\Omega^{1,1}$, both on a compact Calabi--Yau threefold and
locally. Setting $\Psi=\dbar\gamma\in\Omega^{1,2}$ we thus have $\dbar A'=\partial\Psi$, and we define
 \be
  \partial^{-1}\dbar A' := \Psi \;. 
 \ee
The form $\gamma$ is determined up to $\gamma_0$ with $\partial\dbar\gamma_0=0$, which changes $\Psi$ by $\dbar\gamma_0$.
The BCOV action for Kodaira--Spencer gravity for $\partial A'=0$ can then be written as \cite{BCOV} 
\begin{equation}\label{eq:nonlocal}
S[A']=\tfrac12\!\int_X A'\wedge\partial^{-1}\dbar A'
+\tfrac16\!\int_X\omega\wedge\big((A\wedge A\wedge A)\lrcorner\omega\big),
\qquad A=A'\lrcorner\omega^{-1}\;.
\end{equation}
The quadratic term does not depend on the choice of $\gamma$. Indeed, integrating by parts on the
compact manifold $X$, the change $\int_XA'\wedge\dbar\gamma_0$ equals, up to sign,
$\int_X\dbar A'\wedge\gamma_0=\int_X\partial\dbar\gamma\wedge\gamma_0$, which after two further integrations by parts
becomes  $\int_X\gamma\wedge\partial\dbar\gamma_0=0$.
Variation of the action yields 
 \be
 \delta S=\int_X\delta A'\wedge \left(\partial^{-1}\dbar A'-\tfrac12\,(A\wedge A)\lrcorner\omega\right)\;, 
\ee
and acting on the resulting equations of motion with $\partial$ we have 
\be
 \dbar A'-\tfrac12\,\partial\big((A\wedge A)\lrcorner\omega\big)=0\;. 
\ee
Using \eqref{eq:divomega} and $\dv A=0$ it  follows that 
  $\partial\big((A\wedge A)\lrcorner\omega\big) = -\{A, A\}$, 
and so the above equations are  equivalent to the Kodaira--Spencer equations \eqref{eq:KS} in the form 
$\big(\dbar A+\tfrac12\{A,A\}\big)\lrcorner\omega=0$.

\section{Local action for Kodaira--Spencer gravity}\label{sec:local}

In this section we present the local action principle for Kodaira--Spencer gravity formulated as an 
$L_\infty$ algebra. In the first subsection, we introduce 
the Leibniz--Loday algebra of holomorphic one-forms, which closely follows \cite{HS} 
and defines a  Courant algebroid and an 
associated $L_\infty$ algebra due to Roytenberg and Weinstein. 
In the second subsection, we present an $L_\infty$ algebra on a subcomplex of the
Dolbeault bicomplex which encodes the field equations, the gauge symmetries and their algebra.
In the third subsection, we relate this $L_\infty$ algebra to a cyclic one and give the
corresponding BV master action. In the fourth subsection, we reduce the theory to three dimensions.
Throughout this section  we ignore global issues,
as appropriate for instance for $X=\mathbb C^3$, leaving the inclusion of cohomology and 
complex structure moduli to section~\ref{sec:cohom}.

\subsection{Leibniz--Loday algebra and associated $L_\infty$ structure}\label{sec:Linfty}

As outlined in the introduction we now introduce a new `prepotential' field by solving \eqref{A'Closed}: 
 \begin{equation}\label{eq:prepot}
   A'=\partial e 
   \qquad \Leftrightarrow \qquad A_{\bar\jmath}{}^{i} = \omega^{ijk}\partial_j e_{k\bar\jmath}\;, 
 \end{equation}
which in turn introduces a new gauge redundancy in addition to that encoded in \eqref{eq:KSgauge}. 
The corresponding gauge algebra  is governed by a Leibniz--Loday algebra originating 
from an \textit{embedding tensor} map $\vth:\Omega^{1,0}\to\Gamma(T^{1,0})$ from 
holomorphic one-forms to vectors, defined by 
\begin{equation}\label{eq:theta}
(\vth(\alpha))^i=\omega^{ijk}\partial_j\alpha_k
\qquad \Leftrightarrow \qquad \iota_{\vth(\alpha)}\, \omega=\partial\alpha\; .
\end{equation}
These vector fields are divergence-free, 
  \be
  \dv(\vth(\alpha))=0\;. 
  \ee
 Furthermore, the embedding tensor map has a non-trivial kernel given by 
  \be\label{thetakernel} 
    \ker\vth=\partial\, \Omega^{0,0} = \{\partial\alpha=0\}\;. 
  \ee

Next, we define the Leibniz--Loday algebra on this space of $(1,0)$ forms as the bilinear `product' 
\begin{equation}\label{eq:dorfman}
\alpha_1\circ\alpha_2:=\Lie_{\vth(\alpha_1)}\alpha_2 \;,
\end{equation}
where the Lie derivative acts on the holomorphic sector as $\Lie_X=\iota_X\partial+\partial\iota_X$.
Note that this operation is not antisymmetric. Rather, it has a symmetric part, which we will see momentarily 
is encoded in the following bilinear operation, written in the notation of
\cite{Hohm:2018ybo,Bonezzi:2019ygf}:  
\begin{equation}\label{eq:dorfman2}
\alpha_1\bullet\alpha_2:=\iota_{\vth(\alpha_1)}\alpha_2+\iota_{\vth(\alpha_2)}\alpha_1\in\Omega^{0,0}\;.
\end{equation}
The properties of Leibniz--Loday algebras are encoded in the following series of propositions:

\begin{prop}[Quadratic constraint]\label{prop:anchor}
 \be
   \vth(\alpha_1\circ\alpha_2)=[\vth(\alpha_1),\vth(\alpha_2)] \;.
 \ee  
\end{prop}
\begin{proof}
By \eqref{eq:theta} and $\partial\omega=0$ we have  $\Lie_{\vth(\alpha_1)}\omega=\iota_{\vth(\alpha_1)}\partial\omega + \partial(\iota_{\vth(\alpha_1)}\omega) = \partial\partial \alpha_1= 0$. 
Using $\partial\alpha_2=\iota_{\vth(\alpha_2)}\omega$ it  follows that 
$\Lie_{\vth(\alpha_1)}(\partial\alpha_2)=\Lie_{\vth(\alpha_1)}\iota_{\vth(\alpha_2)}\omega
=\iota_{[\vth(\alpha_1),\vth(\alpha_2)]}\omega+\iota_{\vth(\alpha_2)}\Lie_{\vth(\alpha_1)}\omega=\iota_{[\vth(\alpha_1),\vth(\alpha_2)]}\omega$.
Using  $[\partial,\Lie_X]=0$ and \eqref{eq:theta}, we have $\iota_{\vth(\alpha_1\circ \alpha_2)}\omega = \partial(\alpha_1\circ\alpha_2)=\Lie_{\vth(\alpha_1)}\partial\alpha_2
=\iota_{[\vth(\alpha_1),\vth(\alpha_2)]}\omega$. The claim then follows with  nondegeneracy of $\omega$. 
\end{proof}

\begin{prop}[Leibniz--Loday identity]\label{prop:leibniz}
 \be
  \alpha_1\circ(\alpha_2\circ\alpha_3)
=(\alpha_1\circ\alpha_2)\circ\alpha_3+\alpha_2\circ(\alpha_1\circ\alpha_3)\;.
\ee
\end{prop}
\begin{proof}
The left-hand side is $\Lie_{\vth(\alpha_1)}\Lie_{\vth(\alpha_2)}\alpha_3$. By Proposition \ref{prop:anchor},
$(\alpha_1\circ\alpha_2)\circ\alpha_3=\Lie_{[\vth(\alpha_1),\vth(\alpha_2)]}\alpha_3$, so the right-hand  side is
$\Lie_{[\vth(\alpha_1),\vth(\alpha_2)]}\alpha_3+\Lie_{\vth(\alpha_2)}\Lie_{\vth(\alpha_1)}\alpha_3=\Lie_{\vth(\alpha_1)}\Lie_{\vth(\alpha_2)}\alpha_3$ using
$\Lie_{[\vth(\alpha_1),\vth(\alpha_2)]}=[\Lie_{\vth(\alpha_1)},\Lie_{\vth(\alpha_2)}]$.
\end{proof}

\begin{prop}[Symmetric part is trivial]\label{prop:sym}
 \be 
    \alpha_1\circ\alpha_2+\alpha_2\circ\alpha_1
=\partial(\alpha_1\bullet\alpha_2)\in\partial\,\Omega^{0,0}=\ker\vth\;.
\ee
\end{prop}
\begin{proof}
$\Lie_{\vth(\alpha_1)}\alpha_2+\Lie_{\vth(\alpha_2)}\alpha_1
=\iota_{\vth(\alpha_1)}\partial\alpha_2+\iota_{\vth(\alpha_2)}\partial\alpha_1
+\partial(\iota_{\vth(\alpha_1)}\alpha_2+\iota_{\vth(\alpha_2)}\alpha_1)$. The first two cancel:
$\iota_{\vth(\alpha_1)}\iota_{\vth(\alpha_2)}\omega+\iota_{\vth(\alpha_2)}\iota_{\vth(\alpha_1)}\omega=0$. The remainder is
$\partial(\alpha_1\bullet\alpha_2)$ by \eqref{eq:dorfman}.
\end{proof}

We have now established that the above structure exhibits a Courant algebroid structure in 
the sense of \cite{RW}.\footnote{Strictly speaking this is a mild generalization of a Courant algebroid since 
here the anchor is a differential operator and not a bundle map, but it was noted in \cite{HS} that 
the $L_{\infty}$ algebra exists without that assumption.} Indeed, 
with $\vth\circ\partial=0$, cf.~eq.~\eqref{thetakernel}, the  Courant axiom 4 of \cite[Def.~3.2]{RW} 
and $\partial f\bullet\alpha=\rho(\alpha)f$, where here $\rho=\vth$, Propositions
\ref{prop:anchor}--\ref{prop:sym} exhibit $(\Omega^{1,0};\vth,\bullet,
[\cdot,\cdot]_C)$ as the holomorphic Courant algebroid \cite{HS}, with
Courant bracket 
\be 
  [\alpha_1,\alpha_2]_C:=\tfrac12(\alpha_1\circ\alpha_2-\alpha_2\circ\alpha_1)\;.
\ee  
By \cite[Thm.~4.3]{RW} a Courant algebroid yields an $L_\infty$ algebra. Here this $L_\infty$ algebra lives on
the \emph{two-term} complex
\begin{equation}\label{eq:res}
L_{\mathrm{RW}} \ \equiv \ \Big(\; \Omega^{0,0}\ \xrightarrow{\ \partial\ }\ \Omega^{1,0}\;\Big) \;, 
\end{equation}
where the degrees are $\deg\Omega^{0,0}=-1,\  \deg\Omega^{1,0}=0$. 
The $L_\infty$ structure maps are given by $\ell_1=\partial$, together with 
\begin{equation}\label{eq:RWmaps}
\begin{split} 
\ell_2(\alpha_1,\alpha_2)&=[\alpha_1,\alpha_2]_C\,,\\ 
\ell_2(\alpha,f)&=\tfrac12\, \alpha\bullet\partial f = \tfrac{1}{2} \iota_{\vartheta( \alpha)}\partial f \,,\\ 
\ell_3(\alpha_1,\alpha_2,\alpha_3)&=-T(\alpha_1,\alpha_2,\alpha_3) \, ,
\end{split} 
\end{equation}
and Nijenhuis function \cite{RW}
 \be 
    T=   \tfrac12\, [\alpha_{[1},\alpha_2]_C\bullet\alpha_{3]}
    = \tfrac16\big( [\alpha_{1},\alpha_2]_C\bullet\alpha_{3} + \text{two more}\big) \;, 
 \ee   
using antisymmetrizations with combinatorial weight factors ($A_{[1}B_{2]} = \tfrac12 
(A_{1}B_{2}-A_2B_1)$, etc.).  
 All $\ell_{\ge4}$ are zero.\footnote{
Ref.~\cite{RW} states the result on  $\ker\partial\xhookrightarrow{\iota}\Omega^{0,0}
\xrightarrow{\partial}\Omega^{1,0}$. Here
$\ker\partial$ is the locally constant functions, which are isomorphic to $\mathbb C$.  
This space is acted on trivially and carries no bracket. Hence we can drop it.}

The $L_{\infty}$ relations follow by direct computation. 
For instance, the Leibniz relation requires by degree reasons 
 \be
  \ell_1(\ell_2(\alpha, f)) = \ell_2(\ell_1(\alpha), f) + (-1)^{|\alpha|} \ell_2(\alpha, \ell_1(f)) = \ell_2(\alpha, \ell_1(f)) \;. 
 \ee
For the left-hand side we compute 
 \be\label{LHSLeibnizz} 
   \ell_1(\ell_2(\alpha, f)) = \partial (\ell_2(\alpha, f)) = \tfrac{1}{2}\partial  \iota_{\vartheta( \alpha)}\partial f\;, 
 \ee
and for the right-hand side 
 \be
 \begin{split} 
  \ell_2(\alpha, \ell_1(f)) &= [\alpha ,\partial f]_C = \tfrac{1}{2}(\alpha\circ \partial f  - \partial f \circ \alpha) 
  =\tfrac{1}{2} \alpha\circ \partial f = \tfrac{1}{2} {\cal L}_{\vartheta(\alpha)} \partial f \\
  &=  \tfrac{1}{2} \iota_{\vartheta(\alpha)}\partial\partial f +   \tfrac{1}{2}\partial  \iota_{\vartheta(\alpha)}\partial f 
  = \tfrac{1}{2}\partial  \iota_{\vartheta(\alpha)}\partial f \;, 
 \end{split}  
 \ee
which equals \eqref{LHSLeibnizz}.

\subsection{\texorpdfstring{$L_\infty$}{L-infinity} algebra on the bicomplex and the local action}\label{sec:bicx}

We now present the $L_\infty$ algebra that encodes the field equations, gauge symmetries and
gauge algebra of the local theory. Let us first recall the relevant definitions \cite{Hohm:2017pnh}. An $L_\infty$
algebra is a graded vector space $L=\bigoplus_k L_k$ with multilinear maps
$\ell_n:L^{\otimes n}\to L$ of degree $2-n$ that are graded antisymmetric,
\be\label{eq:antisym}
 \ell_n(\dots,x_i,x_{i+1},\dots)=-(-1)^{x_ix_{i+1}}\,\ell_n(\dots,x_{i+1},x_i,\dots)\;,
\ee
where we recall that  in sign exponents $x$ denotes the degree $|x|$. The maps are subject to the $L_{\infty}$ relations
\be\label{eq:Jn}
 \sum_{i+j=n+1}(-1)^{i(j-1)}
 \sum_{\sigma}(-1)^\sigma\,\epsilon(\sigma;x)\,
 \ell_j\big(\ell_i(x_{\sigma(1)},\dots,x_{\sigma(i)}),x_{\sigma(i+1)},\dots,x_{\sigma(n)}\big) = 0 \;,
\ee
with the inner sum running over $(i,n-i)$-unshuffles, where $\epsilon(\sigma;x)$ is the Koszul sign of the
permutation. The first three relations state that $\ell_1$ squares to zero, that $\ell_1$ is a
derivation of $\ell_2$,
\be
 \ell_1\ell_2(x_1,x_2)=\ell_2(\ell_1x_1,x_2)+(-1)^{x_1}\ell_2(x_1,\ell_1x_2)\;,
\ee
and that the Jacobi identity of $\ell_2$ holds up to $\ell_1$-exact terms governed by $\ell_3$,
\be\label{eq:J3}
\begin{split}
 &\ell_2(\ell_2(x_1,x_2),x_3)+(-1)^{(x_1+x_2)x_3}\,\ell_2(\ell_2(x_3,x_1),x_2)+(-1)^{(x_2+x_3)x_1}\,\ell_2(\ell_2(x_2,x_3),x_1)\\
 &\quad=-\ell_1\ell_3(x_1,x_2,x_3)-\ell_3(\ell_1x_1,x_2,x_3)\\
 &\qquad\;-(-1)^{x_1}\ell_3(x_1,\ell_1x_2,x_3)-(-1)^{x_1+x_2}\ell_3(x_1,x_2,\ell_1x_3)\;.
\end{split}
\ee

The $L_\infty$ algebra of the local theory lives on the following eight spaces of the Dolbeault
bicomplex, which are exchanged in pairs by $\Omega^{p,q}\leftrightarrow\Omega^{3-p,3-q}$:
\begin{center}
\begin{tikzpicture}[
  x=1.75cm, y=1.45cm,
  >={Stealth[length=2.6mm,width=2mm]},
  arr/.style ={->,line width=.5pt,shorten >=3pt,shorten <=3pt},
  lab/.style ={above,inner sep=2pt,font=\small,fill=white},
  labb/.style={below,inner sep=2pt,font=\small,fill=white},
  spc/.style ={inner sep=2pt},
  el/.style  ={inner sep=2pt},
]
% spaces
\node[spc] (O00) at (0, 0) {$\Omega^{0,0}$};
\node[spc] (O10) at (1,-1) {$\Omega^{1,0}$};
\node[spc] (O01) at (1, 1) {$\Omega^{0,1}$};
\node[spc] (O11) at (2, 0) {$\Omega^{1,1}$};
\node[spc] (O22) at (4, 0) {$\Omega^{2,2}$};
\node[spc] (O32) at (5,-1) {$\Omega^{3,2}$};
\node[spc] (O23) at (5, 1) {$\Omega^{2,3}$};
\node[spc] (O33) at (6, 0) {$\Omega^{3,3}$};
% left diamond
\draw[arr] (O00) -- node[labb]{$\partial$}      (O10);
\draw[arr] (O00) -- node[lab] {$\dbar$}         (O01);
\draw[arr] (O10) -- node[labb]{$\dbar$}         (O11);
\draw[arr] (O01) -- node[lab] {$\partial$}      (O11);
% bridge
\draw[arr] (O11) -- node[lab] {$\partial\dbar$} (O22);
% right diamond
\draw[arr] (O22) -- node[labb]{$\partial$}      (O32);
\draw[arr] (O22) -- node[lab] {$\dbar$}         (O23);
\draw[arr] (O32) -- node[labb]{$\dbar$}         (O33);
\draw[arr] (O23) -- node[lab] {$\partial$}      (O33);
% typical elements: separate row below, aligned under the columns
\node[el] at (0,-2.4) {$\chi$};
\node[el] at (1,-2.4) {$\begin{pmatrix}\bar\lambda\\\lambda\end{pmatrix}$};
\node[el] at (2,-2.4) {$e$};
\node[el] at (4,-2.4) {$e^\ast$};
\node[el] at (5,-2.4) {$\begin{pmatrix}\lambda^\ast\\\bar\lambda^\ast\end{pmatrix}$};
\node[el] at (6,-2.4) {$\chi^\ast$};
\end{tikzpicture}
\end{center}

The content is the gauge-for-gauge parameter $\chi\in\Omega^{0,0}$, the gauge parameters
$\lambda\in\Omega^{1,0}$ (diffeomorphisms) and $\bar\lambda\in\Omega^{0,1}$ (the redundancy of the prepotential), the
field $e\in\Omega^{1,1}$, and the spaces of `antifields' $e^\ast\in\Omega^{2,2}$, $\lambda^\ast\in\Omega^{2,3}$, $\bar\lambda^\ast\in\Omega^{3,2}$
and $\chi^\ast\in\Omega^{3,3}$. The antifield spaces can be interpreted as the spaces of field equations
($e^\ast$), of Noether identities ($\lambda^\ast$ and $\bar\lambda^\ast$, one for each gauge parameter) and of
Noether-for-Noether identities ($\chi^\ast$). Their degrees are $-1$ for $\chi$, $0$ for $\lambda$
and $\bar\lambda$, $1$ for $e$, $2$ for $e^\ast$, $3$ for $\lambda^\ast$ and $\bar\lambda^\ast$, and $4$ for $\chi^\ast$.
Equivalently, for each space $\Omega^{p,q}$ the degree is $p+q-1$ for the four spaces on the left and $p+q-2$ for  the four
spaces on the right. The differential is
\be\label{eq:l1-folded}
\begin{aligned}
 \ell_1(\chi)&=\begin{pmatrix}\dbar\chi\\ \partial\chi\end{pmatrix}\,, &\qquad
 \ell_1\begin{pmatrix}\bar\lambda\\ \lambda\end{pmatrix}&=\partial\bar\lambda+\dbar\lambda\,, &\qquad
 \ell_1(e)&=\partial\dbar e\,,\\[2pt]
 \ell_1(e^\ast)&=\begin{pmatrix}\dbar e^\ast\\ \partial e^\ast\end{pmatrix}\,, &\qquad
 \ell_1\begin{pmatrix}\lambda^\ast\\ \bar\lambda^\ast\end{pmatrix}&=\partial\lambda^\ast+\dbar\bar\lambda^\ast\,, &\qquad
 \ell_1(\chi^\ast)&=0\,,
\end{aligned}
\ee
where, as in the figure, the upper entries are $\bar\lambda\in\Omega^{0,1}$ and $\lambda^\ast\in\Omega^{2,3}$ and the
lower entries $\lambda\in\Omega^{1,0}$ and $\bar\lambda^\ast\in\Omega^{3,2}$. It squares to zero by \eqref{eq:bicx}. It is first order except on the field,
where it is the second-order operator $\partial\dbar$, thereby bridging the 
bidegrees $(1,1)$ and $(2,2)$.

The higher brackets are as follows. We use the Schouten bracket \eqref{eq:koszul}, the product
\eqref{eq:dorfman2}, and write $c=\bar\lambda^\ast\lrcorner\omega^{-1}\in\Omega^{0,2}$ for the antifield $\bar\lambda^\ast$ with its holomorphic
legs stripped off by $\omega^{-1}$. Operations of section~\ref{sec:Linfty} applied to $e$ or $\bar\lambda$ are understood to
act on the holomorphic indices, at fixed antiholomorphic index. On the gauge parameters the binary
bracket is the Courant bracket of section~\ref{sec:Linfty},
\be\label{eq:Lxixi}
 \ell_2(\lambda_1,\lambda_2)=[\lambda_1,\lambda_2]_C\;,
\ee
and the remaining non-vanishing binary brackets are
\be\label{eq:minl2}
\begin{split}
 \ell_2(\lambda,\chi)&=\tfrac12\,\iota_{\vth(\lambda)}\partial\chi\;,\qquad
 \ell_2(\lambda,\bar\lambda)=\tfrac12\,\Lie_{\vth(\lambda)}\bar\lambda\;,\qquad
 \ell_2(\chi,e)=-\tfrac12\,\iota_{\vth(e)}\partial\chi\;,\\
 \ell_2(\lambda,e)&=[\lambda,e]_C=\Lie_{\vth(\lambda)}e-\tfrac12\,\partial(\lambda\bullet e)\;,\\
 \ell_2(\lambda,y^\ast)&=\Lie_{\vth(\lambda)}y^\ast\quad\text{for}\quad y^\ast\in\{e^\ast,\lambda^\ast,\bar\lambda^\ast,\chi^\ast\}\;,\\
 \ell_2(e_1,e_2)&=-\{\vth(e_1),\vth(e_2)\}\lrcorner\omega\;,\qquad
 \ell_2(e,e^\ast)=\{\vth(e),e^\ast\lrcorner\omega^{-1}\}\lrcorner\omega\;,\\
 \ell_2(e,\bar\lambda^\ast)&=\partial\big((\vth(e)\wedge c)\lrcorner\omega\big)\;,
\end{split}
\ee
where the Lie derivative acts on the holomorphic indices, and $\iota_{\vth(e)}\partial\chi$ is the $(0,1)$-form
obtained by contracting the vector-valued $(0,1)$-form $\vth(e)$ with $\partial\chi$. The ternary brackets are those
of the Roytenberg--Weinstein algebra,
\be\label{eq:minl3}
 \ell_3(\lambda_1,\lambda_2,\lambda_3)=-T(\lambda_1,\lambda_2,\lambda_3)\;,\qquad
 \ell_3(\lambda_1,\lambda_2,e)=-T(\lambda_1,\lambda_2,e)\;\in\;\Omega^{0,1}\;,
\ee
and $\ell_{n\ge4}=0$. All other brackets are either fixed by graded antisymmetry or vanish.

\begin{prop}\label{prop:minimal}
The brackets \eqref{eq:l1-folded}--\eqref{eq:minl3} satisfy all $L_\infty$ relations \eqref{eq:Jn}.
\end{prop}

\noindent
We will call this $L_\infty$ algebra $L$. Its relations follow from the double copy construction of
section~\ref{sec:dc}, in which they are inherited from those of its constituents. We have also
verified them explicitly in components.

The field equation and the gauge transformations follow from the $L_\infty$ algebra in the
standard way. The field equation is the Maurer--Cartan equation\footnote{In the conventions of \cite{Hohm:2017pnh} the Maurer--Cartan equation reads
$\ell_1(e)-\tfrac12\ell_2(e,e)+\dots$. Since our $L_\infty$ algebra has no brackets $\ell_{n\ge4}$, sending
$\ell_2\to-\ell_2$ is consistent with all $L_\infty$ relations, being equivalent to the field
redefinition $e\to-e$, and for later convenience we have included this sign change in the
Maurer--Cartan equation.}
\be\label{eq:EOM}
 E(e):=\ell_1(e)+\tfrac12\ell_2(e,e)=\partial\dbar e-\tfrac12\{\vth(e),\vth(e)\}\lrcorner\omega=0\;,
\ee
and it is covariant under the gauge transformations
\be\label{eq:gauge}
 \delta e=\ell_1\begin{pmatrix}\bar\lambda\\ \lambda\end{pmatrix}+\ell_2(e,\lambda)=\dbar\lambda+\partial\bar\lambda-[\lambda,e]_C\;,
\ee
under which $\delta E=\ell_2(E,\lambda)$. Contracting with $\omega^{-1}$ and using
$(\partial\dbar e)\lrcorner\omega^{-1}=-\dbar(\vth(e))$, one finds $E(e)\lrcorner\omega^{-1}=-\big(\dbar A+\tfrac12\{A,A\}\big)$ with
$A=\vth(e)$, so that \eqref{eq:EOM} is precisely the Kodaira--Spencer equation \eqref{eq:KS}, written
in terms of the prepotential. Similarly, since $\vth$ annihilates the $\partial$-exact term in \eqref{eq:minl2}, the
gauge transformations \eqref{eq:gauge} map under $\vth$ to the gauge transformations \eqref{eq:KSgauge} with
parameter $\xi=-\vth(\lambda)$, i.e.\ to the volume-preserving diffeomorphisms of $A$, while the
$\bar\lambda$-transformations are the redundancy of the prepotential. The algebra of gauge transformations
closes off-shell, with the structure constants $\ell_2(\lambda_1,\lambda_2)$ together with the field-dependent
$\bar\lambda$-component $\ell_3(\lambda_1,\lambda_2,e)$. Moreover, the gauge algebra 
is reducible, with $\chi$ parametrizing the trivial parameters
$(\lambda,\bar\lambda)=(\partial\chi,\dbar\chi)$.

We finally note that 
the Courant bracket in \eqref{eq:gauge} could  be traded for the Lie derivative by the field-dependent
parameter redefinition $\bar\lambda\to\bar\lambda+\tfrac12\,\lambda\bullet e$, in terms of which
$\delta e=\dbar\lambda+\partial\bar\lambda-\Lie_{\vth(\lambda)}e$. On the level of $L_\infty$ algebras this is the isomorphism
$f:L\to L'$, a notion to be reviewed below, with $f_1=1$ and $f_2(\lambda,e)=\tfrac12\,\lambda\bullet e$.
The transformed  $L_{\infty}$ brackets, denoted by a prime, 
include $\ell_2'(\lambda,e)=\Lie_{\vth(\lambda)}e$, $\ell_2'(\lambda,\bar\lambda)=\Lie_{\vth(\lambda)}\bar\lambda$, while $\ell_2'(\chi,e)$ and
$\ell_3'(\lambda_1,\lambda_2,e)$ vanish. The price is a $(0,1)$-component of the gauge bracket,
\be\label{eq:minxixi}
 \ell_2'(\lambda_1,\lambda_2)=[\lambda_1,\lambda_2]_C\;+\;\tfrac12\big(\iota_{\vth(\lambda_2)}\dbar\lambda_1-\iota_{\vth(\lambda_1)}\dbar\lambda_2\big)
 -\tfrac12\big(\iota_{\dbar \vth(\lambda_2)}\lambda_1-\iota_{\dbar \vth(\lambda_1)}\lambda_2\big)\;,
\ee
where the second and third terms are $(0,1)$-forms, with the contractions acting on the holomorphic
indices.

\subsection{Cyclic \texorpdfstring{$L_\infty$}{L-infinity} algebra and BV master action}\label{sec:BV}

In order to derive the theory from an action we need, in addition, an inner product that is
compatible with the brackets. The folded complex carries the inner product
\be\label{eq:folded-pairing}
 \cip{a}{b}=\cip{b}{a}:=\int_X b\wedge a\;,
\ee
for $a$ one of $\chi,\lambda,\bar\lambda,e$ and $b$ the corresponding one of $\chi^\ast,\lambda^\ast,\bar\lambda^\ast,e^\ast$, the
antifield being placed first independently of the order of the arguments. It has degree $-3$
and is graded symmetric.\footnote{The ordering matters: since $a\wedge b=(-1)^{k}\,b\wedge a$ for $a$ of form
degree $k$, the literal $\int_Xa\wedge b$ is antisymmetric on the $(\lambda,\lambda^\ast)$ and $(\bar\lambda,\bar\lambda^\ast)$
sectors and hence not an inner product there.} An $L_\infty$ algebra is cyclic with respect to
$\cip{\cdot}{\cdot}$ if
\be\label{eq:cycsign}
 \cip{x_0}{\ell_n(x_1,\dots,x_n)}
 =(-1)^{\,n+n(x_0+x_n)+x_n(x_0+\cdots+x_{n-1})}\,\cip{x_n}{\ell_n(x_0,\dots,x_{n-1})}\;, 
\ee
in the conventions of \cite{JRSW}.
 Together with the graded antisymmetry \eqref{eq:antisym} of the $\ell_n$,
this determines the behavior of $\cip{x_0}{\ell_n(x_1,\dots,x_n)}$ under all permutations of its $n+1$
arguments. The  form is graded symmetric
with respect to the shifted degrees $|x|+1$, corresponding to the suspended so-called $b$-picture, which in turn explains 
 the additional terms in the exponent of \eqref{eq:cycsign}.
For $n=1$, for instance, \eqref{eq:cycsign} states that $\ell_1$ is graded skew-adjoint,
$\cip{\ell_1x_0}{x_1}=-(-1)^{x_0}\cip{x_0}{\ell_1x_1}$. For fields $e$ in degree one we have 
$\cip{\ell_1e}{e}=\cip{e}{\ell_1e}$, as expected. 
More generally, cyclicity follows  for $\ell_1$ defined in  \eqref{eq:l1-folded}  by
integration by parts, using \eqref{eq:bicx}. The higher brackets of
$L$, however, are not cyclic. For instance, cyclicity relates $\cip{\chi^\ast}{\ell_2(\lambda,\chi)}$ to
$\cip{\chi}{\ell_2(\lambda,\chi^\ast)}$, and after integration by parts the factor $\tfrac12$ in
$\ell_2(\lambda,\chi)=\tfrac12\,\iota_{\vth(\lambda)}\partial\chi$ is not matched by $\ell_2(\lambda,\chi^\ast)=\Lie_{\vth(\lambda)}\chi^\ast$. The algebra $L$ is, however, isomorphic to a cyclic one.

To show this we have to  recall the notion of  $L_\infty$ morphisms. These are maps  $F:L\to L'$ consisting  of graded antisymmetric multilinear
maps $F_n:L^{\otimes n}\to L'$ of degree $1-n$, subject to relations that ensure compatibility with the
brackets. In the lowest arities they state that $F_1$ is a chain map, $\ell'_1F_1=F_1\ell_1$, and that
$F_1$ intertwines the binary brackets up to a homotopy given by $F_2$,
\be\label{eq:morph2}
 \ell'_2(F_1x_1,F_1x_2)=F_1\ell_2(x_1,x_2)+\ell'_1F_2(x_1,x_2)+F_2(\ell_1x_1,x_2)+(-1)^{x_1}F_2(x_1,\ell_1x_2)\;.
\ee
The relation in arity three similarly expresses $\ell'_3(F_1x_1,F_1x_2,F_1x_3)$ in terms of $F_1\ell_3$,
of $F_3$ composed with $\ell_1$, and of compositions of $F_2$ with the binary brackets. A morphism
with $F_{n\ge2}=0$ is called strict, and a morphism with invertible $F_1$ is an isomorphism. Given an
$L_\infty$ algebra $L'$ and maps $F_n$ with invertible $F_1$, the morphism relations can be read as
definitions of brackets $\ell_n$ on the underlying complex of $L$, arity by arity, and these brackets
then satisfy the $L_\infty$ relations because those of $L'$ do.

\begin{prop}\label{prop:master}
Let $F_1=1$, $F_{n\ge3}=0$, and let the only non-vanishing components of $F_2$ be
\be\label{eq:iso}
\begin{split}
 F_2(\lambda,e)&=-\tfrac12\,\lambda\bullet e\;,\\
 F_2(\lambda,\chi^\ast)&=\partial\lambda\wedge c_\chi-\tfrac12\,\lambda\wedge\partial c_\chi\;,\qquad
 F_2(e,\bar\lambda^\ast)=2\,\partial e\wedge c+e\wedge\partial c\;,
\end{split}
\ee
where $c=\bar\lambda^\ast\lrcorner\omega^{-1}\in\Omega^{0,2}$ and $c_\chi=\chi^\ast\lrcorner\omega^{-1}\in\Omega^{0,3}$. The $L_\infty$ algebra
$L_{\rm cyc}$ on the folded complex for which $F:L_{\rm cyc}\to L$ is an $L_\infty$ isomorphism is
cyclic with respect to \eqref{eq:folded-pairing}.
\end{prop}

\noindent
Explicitly, \eqref{eq:morph2} gives the binary brackets of $L_{\rm cyc}$ as
\be
 \ell_2^{\rm cyc}(x_1,x_2)=\ell_2(x_1,x_2)-\ell_1F_2(x_1,x_2)-F_2(\ell_1x_1,x_2)-(-1)^{x_1}F_2(x_1,\ell_1x_2)\;,
\ee
and the ternary bracket follows in the same way from the relation in arity three, while
$\ell^{\rm cyc}_{n\ge4}=0$. By construction, $L_{\rm cyc}$ is an $L_\infty$ algebra isomorphic to $L$: its brackets are
defined by the morphism relations, as explained above, and they satisfy the $L_\infty$ relations because
those of $L$ do. The non-trivial statement of Proposition \ref{prop:master} is that these brackets are cyclic.
We have verified this by direct computation, i.e.\ by checking \eqref{eq:cycsign} for $\ell^{\rm cyc}_2$ and $\ell^{\rm cyc}_3$ on all
combinations of fields. The map \eqref{eq:iso} itself can be found in the same way: making the most
general ansatz for $F_2$ compatible with the degrees and with conservation of the antiholomorphic degree,
the conditions \eqref{eq:cycsign} for the transported brackets are linear in the coefficients of the ansatz,
and solving them gives \eqref{eq:iso}. No $F_3$ is possible by the same degree counting.

The component $F_2(\lambda,e)$ is the field-dependent redefinition of $\bar\lambda$ discussed at the end of
section~\ref{sec:bicx}: composing $F$ with the isomorphism $f:L\to L'$ removes it, so that $L_{\rm cyc}$ is related to $L'$ by
an isomorphism whose only components are the two $\lambda^\ast$-valued ones in \eqref{eq:iso}. Consequently,
$\ell^{\rm cyc}_2(\lambda,e)=\Lie_{\vth(\lambda)}e$, as in the master action below, and $\ell^{\rm cyc}_2(\lambda_1,\lambda_2)=\ell'_2(\lambda_1,\lambda_2)$, including the
$(0,1)$-component in \eqref{eq:minxixi}. The other two components of $F_2$ take values in the $\lambda^\ast$-sector. They account for a factor
$\tfrac12$ in the Lie derivative on $\chi^\ast$, which repairs the example above, for additional
brackets $\ell_2(\chi,\chi^\ast)$, $\ell_2(\bar\lambda,\bar\lambda^\ast)$ and $\ell_2(\lambda,\bar\lambda^\ast)$ with values in $\Omega^{2,3}$, for the
different form of $\ell_2(\lambda,\lambda^\ast)$ and $\ell_2(e,e^\ast)$, for the absence of $\ell_2(e,\bar\lambda^\ast)$, and for the
cyclic partner $\ell_3(\lambda,\lambda,\chi^\ast)\in\Omega^{2,3}$ of $\ell_3(\lambda,\lambda,\lambda)$.

Instead of listing  the cyclic $L_{\infty}$ brackets explicitly we encode them in the associated BV master action. 
Given a cyclic $L_\infty$ algebra, the corresponding BV action is
\be\label{eq:bracketsfromS}
 S[\Phi]=\sum_{n\ge1}\frac{1}{(n+1)!}\,\cip{\Phi}{\ell_n(\Phi,\dots,\Phi)}\;,
\ee
where $\Phi$ is the BV superfield of total degree one. Concretely,
\be\label{eq:superfield}
 \Phi=\chi+\lambda+\bar\lambda+e+e^\ast+\lambda^\ast+\bar\lambda^\ast+\chi^\ast\;,
\ee
where each component is a form of the bidegree given in section~\ref{sec:bicx}, but with coefficient
functions taking values in a Grassmann algebra, for instance $\lambda=\lambda_i(z,\bar z)\,dz^i$ with
anticommuting $\lambda_i$. The coefficients carry the ghost number, defined as one minus the $L_\infty$
degree: $2$ for $\chi$, $1$ for $\lambda$ and $\bar\lambda$, $0$ for $e$, and $-1,-2,-2,-3$ for
$e^\ast,\lambda^\ast,\bar\lambda^\ast,\chi^\ast$, and their Grassmann parity is the ghost number mod $2$. In
finite dimensions this corresponds to writing $\Phi=\sum_a\tau_a\phi^a$ with a basis $\tau_a$ and graded
coordinates $\phi^a$. Here the role of the label $a$ is played by the form indices and the point of
$X$, and the sum is the component expansion of the forms. The brackets and the inner product are
extended to such Grassmann-valued forms by multilinearity, with a Koszul sign whenever a
Grassmann-odd coefficient is moved past a form leg or past another odd coefficient. Since $\Phi$ has
total degree one, the graded antisymmetry \eqref{eq:antisym} turns into ordinary symmetry when all
arguments of $\ell_n$ equal $\Phi$, so that $\ell_n(\Phi,\dots,\Phi)$ is in general non-zero. Expanding
\eqref{eq:bracketsfromS} in components then gives the action, the factor $1/(n+1)!$ cancelling against
the multiplicities of the terms. Conversely, the action determines the brackets: by cyclicity
$\cip{x_0}{\ell_n(x_1,\dots,x_n)}$ is obtained from the term of order $n+1$ in $S$ by polarization. 
The $L_\infty$ relations of a cyclic $L_\infty$ algebra are
equivalent to the classical master equation of $S$. By Propositions \ref{prop:minimal}
and \ref{prop:master}, the action of $L_{\rm cyc}$ therefore solves the classical master equation. In
components it reads
\be\label{eq:master}
\begin{aligned}
 S=\int_X\Big[\;
 &\tfrac12\,e\wedge\partial\dbar e
 -e^\ast\wedge\big(\dbar\lambda+\partial\bar\lambda-\Lie_{\vth(\lambda)}e\big)
 +\lambda^\ast\wedge\big(\partial\chi+\tfrac12[\lambda,\lambda]_C\big)\\
 &+\bar\lambda^\ast\wedge\big(\dbar\chi+\Lie_{\vth(\lambda)}\bar\lambda+\tfrac12\,\ell'_2(\lambda,\lambda)\big|_{\Omega^{0,1}}\big)
 +\chi^\ast\wedge\big(\tfrac12\,\iota_{\vth(\lambda)}\partial\chi+\tfrac16\,T(\lambda,\lambda,\lambda)\big)\Big]\\
 &+\tfrac16\int_X\omega\wedge\big((A\wedge A\wedge A)\lrcorner\omega\big)\;,
\end{aligned}
\ee
where the Grassmann parameters are ordered as written, $A=\vth(e)$ is the Beltrami differential of
section~\ref{sec:bicx}, and $\ell'_2(\lambda,\lambda)|_{\Omega^{0,1}}$ is the $(0,1)$-form part of the bracket \eqref{eq:minxixi}.
With the conventions of section~\ref{sec:review} one has, up to a total derivative,
$\int_X\omega\wedge((A\wedge A\wedge A)\lrcorner\omega)=-\int_Xe\wedge\{\vth(e),\vth(e)\}\lrcorner\omega$. The structure is
the familiar one: the classical action
$\tfrac12\int e\wedge\partial\dbar e+\tfrac16\int\omega\wedge((A\wedge A\wedge A)\lrcorner\omega)$, which is precisely the
local action, followed by the antifields multiplying the
gauge transformations of the corresponding fields, in the Lie derivative form
$\delta e=\dbar\lambda+\partial\bar\lambda-\Lie_{\vth(\lambda)}e$ discussed at the end of section~\ref{sec:bicx}, and the terms
encoding the reducibility.

\subsection{Reduction to three dimensions}\label{sec:3d}

We now reduce  the theory to the real slice of $X=\mathbb C^3$ defined by $z^i=\bar{z}^{\bar\imath}$. 
The fields are thus independent of the imaginary parts $y^i$ in \eqref{DEFzandbarz}. Furthermore, 
we rescale the real parts of the coordinates  by a factor of $2$ so that $x^i=z^i+\bar z^i$, and 
\be\label{eq:slice}
 \partial_i=\dbar_{\bar\imath} = \frac{\partial}{\partial x^i}
 \;.
\ee
Holomorphic and antiholomorphic indices then both become indices of $\mathbb R^3$, 
so that $e_{i\bar\jmath}$ becomes a
$3\times3$ matrix $e_{ij}$. Moreover, the 3-form $\omega_{ijk}$ reduces to the Levi-Civita symbol $\epsilon_{ijk}$ of $\mathbb R^3$. 
We normalize the volume factor of the imaginary parts $y^i$ to one, i.e.\
$\int_Xf\,dz^1\wedge dz^2\wedge dz^3\wedge d\bar z^1\wedge d\bar z^2\wedge d\bar z^3=\int d^3x\,f$, and from now on write $\partial_i$ for $\partial/\partial x^i$.
The classical action $\tfrac12\int e\wedge\partial\dbar e+\tfrac16\int\omega\wedge((A\wedge A\wedge A)\lrcorner\omega)$ then reduces to
\be\label{eq:S3d}
 S=-\int d^3x\,\Big(\tfrac12\,\epsilon^{ikm}\epsilon^{jln}\,e_{ij}\,\partial_k\partial_l\,e_{mn}+\det A\Big)\;,\qquad
 A^i{}_j=\epsilon^{ikl}\partial_ke_{lj}\;,
\ee
where $A$ is the Beltrami differential $\vth(e)$. The gauge transformations in \eqref{eq:master} become
\be\label{eq:gauge3d}
 \delta e_{ij}=\partial_i\bar\lambda_j-\partial_j\lambda_i-\big(X^k\partial_ke_{ij}+\partial_iX^k\,e_{kj}\big)\;,\qquad
 X^i=\vth(\lambda)^i= \epsilon^{ijk}\partial_j\lambda_k\;.
\ee
The field-dependent term is the Lie derivative along the divergence-free vector field $X$, acting on the
first index of $e$. Note that  with $\partial_kX^k=0$ one computes 
\be\label{eq:Jacobi3d}
 \delta\det A=-\partial_k\big(X^k\det A\big)\;, 
\ee
which in turn implies that the cubic term is invariant under the field-dependent transformations, up to a total derivative,
for arbitrary $e$. In fact,  \eqref{eq:S3d} is exactly gauge  invariant under \eqref{eq:gauge3d}, as expected from 
the gauge invariance of the full Kodaira--Spencer theory. 

We next decompose $e$ into its symmetric and antisymmetric parts,
\be\label{fieldSplit} 
 e_{ij}=h_{ij}+b_{ij}\;,\qquad h_{ij}=h_{(ij)}\;,\qquad b_{ij}=:\epsilon_{ijk}v^k\;.
\ee
The operator $\epsilon^{ikm}\epsilon^{jln}\partial_k\partial_l$ is invariant under the simultaneous exchange $i\leftrightarrow j$, $m\leftrightarrow n$, so that
the terms mixing $h$ and $b$ in the quadratic action are total derivatives. Using that the  linearized Einstein tensor $G^{ij}$
in three dimensions can be written as \cite{Bergshoeff:2009hq}
 \be\label{eq:G3d}
   G^{ij}(h) = \tfrac{1}{2}\, \epsilon^{ikm}\epsilon^{jln}\partial_k\partial_lh_{mn} \,, 
 \ee  
and introducing the field strength $H=db$, whose only independent component is $H_{123}=\partial_kv^k$, we obtain
\be\label{eq:FPKR}
\begin{split} 
 S_2&= -\int d^3x\;h^{ij}\,G_{ij}(h)+\int d^3x\;(\partial_kv^k)^2 \\
 &=-\int d^3x\;h^{ij}\,G_{ij}(h)+\tfrac16\int d^3x\;H_{ijk}H^{ijk}\;.
\end{split} 
\ee
The first term is the Fierz--Pauli action, twice the quadratic part of $\int\sqrt g\,R$ for $g_{ij}=\delta_{ij}+h_{ij}$. 
The second term is the kinetic term of a Kalb--Ramond field, which relative to the combination $R-\tfrac1{12}H^2$
of non-topological string theory has the opposite sign, but we are in Euclidean signature, and a two-form is anyhow not 
propagating in three dimensions. The linearized gauge
transformations read
\be\label{eq:DFTgauge}
\begin{split}
 \delta h_{ij}&=2\,\partial_{(i}\xi_{j)}\;,\qquad\;\, \xi=\tfrac12(\bar\lambda-\lambda)\;,\\
 \delta b_{ij}&=2\,\partial_{[i}\Lambda_{j]}\;,\qquad\, \Lambda=\tfrac12(\bar\lambda+\lambda)\;,
\end{split}
\ee
which are the linearized diffeomorphisms and $B$-field gauge transformations, with $\lambda$ and $\bar\lambda$ entering precisely as 
in  double field theory \cite{Hull:2009mi}. In contrast to standard double field theory, no dilaton is needed.  
On-shell, the $b$ field just gives an integration constant: varying the action with respect to $v^k$ yields 
$\partial_k(\partial_l v^l)=0$, which implies $v^k=m x^k$, with $m$ constant, up to a  pure gauge part $\delta v^k =\epsilon^{klm}\partial_l\Lambda_m$  
that we set to zero.

Let us now turn to the cubic interactions. Remarkably, they are not those of  Einstein--Hilbert gravity, because 
the vertex has three derivatives and breaks parity. Moreover, 
the gauge transformations act through the divergence-free vector fields $X$ rather than through the
parameter $\xi$ of \eqref{eq:DFTgauge}. Such deformations are not covered by uniqueness results as given in 
\cite{BDGH}, which assume parity invariance and at most two derivatives. However, a deformation of the type found here was
constructed  by Boulanger and Gualtieri \cite{BG}. To compare with \cite{BG}, we set $b=0$. Then
$-\det A=4\,a_0$, with
\be
 a_0=\tfrac13\,\epsilon^{ijk}\delta^{lm}\,\partial_{[l}h_{n]i}\,\partial_{[n}h_{p]j}\,\partial_{[m}h_{p]k}\;, 
\ee
which is precisely the vertex  given in eq.~(1.3) of \cite{BG}.

To analyze  the consistency of the theory beyond cubic order, let us inspect the full gauge transformations for non-vanishing $h$ and $b$.
Inserting $\lambda=\Lambda-\xi$ and $\bar\lambda=\Lambda+\xi$ into \eqref{eq:gauge3d} and decomposing into symmetric and antisymmetric parts,
we find
\be\label{eq:gaugehb}
\begin{split}
 \delta h_{ij}&=2\,\partial_{(i}\xi_{j)}-X^k\partial_kh_{ij}-\partial_{(i}X^k\,h_{j)k}-\partial_{(i}X^k\,b_{|k|j)}\;,\\
 \delta b_{ij}&=2\,\partial_{[i}\Lambda_{j]}-X^k\partial_kb_{ij}-\partial_{[i}X^k\,h_{j]k}-\partial_{[i}X^k\,b_{|k|j]}\;,
\end{split}
\ee
with $X^i=\epsilon^{ijk}\partial_j(\Lambda_k-\xi_k)$ and (anti-)symmetrization with unit weight. Setting $b=0$ is not consistent
with these gauge transformations, because the transformation of $b$ contains the term $-\partial_{[i}X^k\,h_{j]k}$, 
and so a $b$ is necessarily generated, even if initially it is set to zero. 
(Moreover, this term cannot be removed by a parameter redefinition: for general $h$ it is not of the form 
$\partial_{[i}\Lambda'_{j]}$.)  Up to
cubic order, however, it is easy to see that the truncation is consistent. 
For $b=0$ and $\Lambda=0$, \eqref{eq:gaugehb} reduces to
\be\label{eq:gaugeBG}
 \delta h_{ij}=2\,\partial_{(i}\xi_{j)}-X^k\partial_kh_{ij}-\partial_{(i}X^k\,h_{j)k}\;,\qquad X^i=-\epsilon^{ijk}\partial_j\xi_k\;,
\ee
which agrees with  eq.~(4.37) in \cite{BG}, up to a field-dependent redefinition of
$\xi$ and the normalization of the coupling. Beyond cubic order, the truncated  theory
is not gauge invariant: the variation of its cubic vertex under the field-dependent part of
\eqref{eq:gaugeBG} does not vanish, not even up to equations of motion and total derivatives.

Let us finally sketch how this infinite series is related to integrating out $b$. In terms of $h$ and $b$ or $v$, cf.~\eqref{fieldSplit}, 
the Beltrami differential reads
\be\label{eq:Ahv}
 A^i{}_j=M^i{}_j+\delta^i_j\,\partial_kv^k-\partial_jv^i\;,\qquad M^i{}_j=\epsilon^{ikl}\partial_kh_{lj}\;. 
\ee
Expanding the vertex in powers of $v$, we obtain 
\be
 \det A=\det M+C^j{}_i\big(\delta^i_j\,\partial_kv^k-\partial_jv^i\big)+\mathcal O(v^2)\;,\qquad
 C^j{}_i=\frac{\partial\det M}{\partial M^i{}_j}=\tfrac12\,\epsilon_{ikl}\epsilon^{jmn}M^k{}_mM^l{}_n\;,
\ee
where we defined the cofactor matrix $C$ of $M$, which obeys $C^j{}_iM^i{}_k=\det M\,\delta^j_k$ and is quadratic in $h$. Up to total
derivatives, the cubic term linear in $v$ is 
\be\label{eq:Jcurrent}
 S_3\big|_{{\cal O}(v)} =- \int d^3x\, v^iJ_i\;, \quad {\rm where} \quad J_i = \partial_jC^j{}_i-\partial_iC^j{}_j\;.
\ee
In the free theory, $\partial_kv^k=3m$ is constant and the transverse part of $v$ is pure gauge, so that integrating
out $v$ leaves the Fierz--Pauli action plus a constant. At first order in the coupling, $v$ takes its free
value, and for $m=0$ we obtain the Boulanger--Gualtieri vertex. (A non-zero $m$ contributes $-2m\int C^i{}_i$, which one may verify to be proportional to the Fierz--Pauli action and hence merely rescales it.) At second order, 
we obtain the field equation $2\,\partial_i(\partial_kv^k)=-J_i$, with $J$ defined in \eqref{eq:Jcurrent}. 
We solve for $v$, upon fixing the gauge for its transverse
part, which yields $\partial_kv^k=-\tfrac12\,\Box^{-1}\partial^iJ_i$ for $m=0$. 
Next, we eliminate $v^i$ in the action terms  \eqref{eq:FPKR} and \eqref{eq:Jcurrent}, where for the latter 
we can contract  $2\,\partial_i(\partial_kv^k)=-J_i$ with $v^i$, to obtain   the quartic term
\be\label{eq:S4}
 S_4=-\tfrac14\int d^3x\,\big(\Box^{-1}\partial^iJ_i\big)^2\;. 
\ee

The above  action is non-local, but its non-localities can be removed by field redefinitions as follows. 
We first note with \eqref{eq:G3d}  that the divergence of the cofactor
matrix is proportional to the free field equation:
\be\label{eq:divC}
 \partial_jC^j{}_i=\epsilon_{ikl}\,\epsilon^{jmn}\,\partial_jM^k{}_m\,M^l{}_n=2\,\epsilon_{ikl}\,G^{kn}(h)\,M^l{}_n\;. 
\ee
The current \eqref{eq:Jcurrent} thus takes the form
$J_i=-\partial_iC^j{}_j+2\,\epsilon_{ikl}\,G^{kn}(h)\,M^l{}_n$,\footnote{Therefore, on-shell $J_i$ is a 
gradient, as needed for the integrability of $2\,\partial_i(\partial_kv^k)=-J_i$. 
} 
and the quartic term \eqref{eq:S4} becomes
\be\label{eq:S4local}
 S_4=-\tfrac14\int d^3x\,\big(C^j{}_j\big)^2+\int d^3x\,C^j{}_j\,N-\int d^3x\,N^2\;,\qquad
 N=\Box^{-1}\partial^i\big(\epsilon_{ikl}\,G^{kn}(h)\,M^l{}_n\big)\;.
\ee
The first term is local. The second and third term are non-local but can be written 
as $\int d^3x\,G^{kn}(h)\,Y_{kn}$, proportional to the free field equation, and hence be removed 
by a field redefinition $h_{kn}\to h_{kn}+\tfrac12\,Y_{kn}$. 
These field redefinitions are non-local and generate further terms at higher order. Up to such
redefinitions, integrating out $b$ therefore yields, to second order in the coupling, the truncated
theory $S|_{b=0}$ together with the local quartic term $-\tfrac14\int d^3x\,(C^j{}_j)^2$. 
Integrating out $b$ at higher orders will  produce an 
infinite series, and it would be interesting to explore whether this reproduces  the second-order formulation of \cite{BG}
to all orders.

\section{Double copy interpretation}\label{sec:dc}

In this section we show that the $L_\infty$ algebra of section~\ref{sec:bicx} is built, up to an explicit
isomorphism, from two `single copies': the Dolbeault algebra of antiholomorphic forms on the one hand, and 
the Roytenberg--Weinstein $L_\infty$ algebra on holomorphic forms on the other, 
with the latter being mapped  to the Lie algebra of 
holomorphic divergence-free vector fields by the embedding tensor. All $L_\infty$ relations then follow from the relations of
the single copies.

\subsection{The single copies}\label{sec:dc-single}

The first single copy is the Dolbeault algebra
\be
 \Kk=\big(\Omega^{0,\bullet},\dbar,\wedge\big)\;,
\ee
with $\Omega^{0,q}$ in degree $q$, which is a differential graded commutative  algebra. The subspace
\be
 \Kk^+:=\Omega^{0,2}\oplus\Omega^{0,3}
\ee
is a differential ideal: it is mapped to itself by $\dbar$ and by multiplication with arbitrary
elements of $\Kk$. We write $\Kk^-=\Omega^{0,0}\oplus\Omega^{0,1}$, so that $\Kk=\Kk^-\oplus\Kk^+$ as a vector
space, and we have the isomorphism $\Kk/\Kk^+\cong\Kk^-$.

The second single copy is the Roytenberg--Weinstein algebra $L_{\rm RW}$ of \eqref{eq:res}, with
brackets \eqref{eq:RWmaps}. For the following construction it is important to display
the dg Lie algebra into which $L_{\rm RW}$ is mapped via the embedding tensor. It is most familiar as
the dg Lie algebra of holomorphic vector fields and functions, $\Gamma(T^{1,0})\xrightarrow{\dv}\mathcal O$, with
the vector fields in degree $0$, the functions in degree $1$, the divergence $\dv X=\partial_iX^i$ as
differential, and the brackets
\be\label{eq:Hvec}
 [X,Y]=\big(X^j\partial_jY^i-Y^j\partial_jX^i\big)\partial_i\;,\qquad [X,f]=X^i\partial_if\;,\qquad [f,g]=0\;.
\ee
For our purposes it is convenient to transport this algebra  to forms with the help of $\omega$, using the
isomorphisms $X\mapsto\iota_X\omega\in\Omega^{2,0}$ and $f\mapsto f\omega\in\Omega^{3,0}$. Since $\partial(\iota_X\omega)=(\dv X)\,\omega$ by
\eqref{eq:divomega}, the divergence becomes the holomorphic differential, and we define
\be\label{eq:Hdgla}
 \mathfrak H=\big(\Omega^{2,0}\xrightarrow{\ \partial\ }\Omega^{3,0}\big)\;,
\ee
with holomorphic $(2,0)$-forms in degree $0$ and holomorphic $(3,0)$-forms in degree $1$. Writing
$\beta=\iota_{X_\beta}\omega$ for $\beta\in\Omega^{2,0}$ and $\gamma=f_\gamma\,\omega$ for $\gamma\in\Omega^{3,0}$, the brackets are
\be\label{eq:Hbrackets}
 [\beta_1,\beta_2]=\iota_{[X_{\beta_1},X_{\beta_2}]}\,\omega\;,\qquad [\beta,\gamma]=X_\beta(f_\gamma)\,\omega\;,\qquad [\gamma_1,\gamma_2]=0\;.
\ee
Up to a sign, the bracket on $\Omega^{2,0}$ is the Schouten bracket \eqref{eq:koszul} transported by $\omega$.

\begin{prop}
$\mathfrak H$ is a dg Lie algebra.
\end{prop}

\begin{proof}
Since \eqref{eq:Hdgla} with \eqref{eq:Hbrackets} is isomorphic to the complex of vector fields and
functions with \eqref{eq:Hvec}, it suffices to verify the claim for the latter. The Jacobi identity for
three vector fields is that of the Lie bracket, and for two vector fields and a function it states
that functions form a representation. That the divergence is a derivation,
$\dv[X,Y]=X(\dv Y)-Y(\dv X)$, follows since the terms $\partial_iX^j\partial_jY^i-\partial_iY^j\partial_jX^i$ cancel. The
derivation property on a vector field and a function holds trivially, since there is nothing in
degree two.
\end{proof}

\noindent
Since $\iota_{\vth(\alpha)}\omega=\partial\alpha$ for $\alpha\in\Omega^{1,0}$ by \eqref{eq:theta}, the embedding tensor $\vth$ becomes, in this language,
simply the holomorphic differential $\partial:\Omega^{1,0}\to\Omega^{2,0}$. The relation between the $L_\infty$ algebra
$L_{\rm RW}$ and the dg Lie algebra $\mathfrak H$ is therefore encoded in the following diagram, which is the
holomorphic de Rham complex split into two halves and then folded:

\begin{center}
\begin{tikzpicture}[
  x=2.6cm, y=1.6cm,
  >={Stealth[length=2.6mm,width=2mm]},
  arr/.style ={->,line width=.5pt,shorten >=3pt,shorten <=3pt},
  lab/.style ={above,inner sep=2pt,font=\small},
  side/.style={right,inner sep=2pt,font=\small},
]
% degrees
\node[font=\small] at (0,-1.75) {$-1$};
\node[font=\small] at (1,-1.75) {$0$};
\node[font=\small] at (2,-1.75) {$1$};
% top row: \mathfrak H, extended by 0 to the left
\node (H0) at (0,0) {$0$};
\node (H1) at (1,0) {$\Omega^{2,0}$};
\node (H2) at (2,0) {$\Omega^{3,0}$};
\node[anchor=west] at (2.35,0) {$\mathfrak H$};
% bottom row: L_RW, extended by 0 to the right
\node (L0) at (0,-1) {$\Omega^{0,0}$};
\node (L1) at (1,-1) {$\Omega^{1,0}$};
\node (L2) at (2,-1) {$0$};
\node[anchor=west] at (2.35,-1) {$L_{\rm RW}$};
% horizontal maps
\draw[arr] (H0) -- (H1);
\draw[arr] (H1) -- node[lab]{$\partial$} (H2);
\draw[arr] (L0) -- node[lab]{$\partial$} (L1);
\draw[arr] (L1) -- (L2);
% vertical maps: the embedding tensor \Pi
\draw[arr] (L0) -- node[side]{$0$} (H0);
\draw[arr] (L1) -- node[side]{$\partial$} (H1);
\draw[arr] (L2) -- node[side]{$0$} (H2);
\end{tikzpicture}
\end{center}

\begin{prop}\label{prop:anchor-strict}
The map $\Pi:L_{\rm RW}\to\mathfrak H$ given by $\Pi(f)=0$ and $\Pi(\alpha)=\partial\alpha$ is a strict $L_\infty$ morphism.
\end{prop}

\begin{proof}
$\Pi$ is a chain map by $\partial^2=0$. Furthermore, $\Pi$
intertwines the binary brackets: since $X_{\partial\alpha}=\vth(\alpha)$, the relation $\partial[\alpha_1,\alpha_2]_C=[\partial\alpha_1,\partial\alpha_2]$ is
equivalent to $\vth([\alpha_1,\alpha_2]_C)=[\vth(\alpha_1),\vth(\alpha_2)]$, which holds by Proposition \ref{prop:anchor}, and
$\Pi\,\ell_2(\alpha,f)=0=[\partial\alpha,\Pi f]$. Finally, $\Pi\,\ell_3=0$ since $\ell_3$ takes values in $\Omega^{0,0}=\ker\Pi$, while
$\mathfrak H$ has no ternary bracket.
\end{proof}

Note that the non-strictness of $L_{\rm RW}$ is invisible to $\Pi$, precisely because the image of its
ternary bracket lies in the kernel of $\Pi$.

\subsection{The double copy}\label{sec:dc-fold}

Let us first recall how tensor products produce $L_\infty$ algebras. Let $\mathcal A$ be a graded
commutative associative differential graded algebra, with differential $d$ and product written as juxtaposition,
$a_1a_2\cdots a_n$, and let $\mathfrak L$ be an $L_\infty$ algebra with brackets $\ell_n$. Then
$\mathcal A\otimes\mathfrak L$ is an $L_\infty$ algebra with brackets
\be\label{eq:tensorprinciple}
\begin{split}
 \widehat{\ell}_1(a\otimes x)&:=da\otimes x+(-1)^{a}\,a\otimes\ell_1(x)\;,\\
 \widehat{\ell}_n(a_1\otimes x_1,\dots,a_n\otimes x_n)&:=\pm\,a_1\cdots a_n\otimes\ell_n(x_1,\dots,x_n)\;,\qquad n\ge2\;,
\end{split}
\ee
where the sign is the Koszul sign of moving $\ell_n$ and the $x_i$ past the $a_j$. Its $L_\infty$
relations follow from those of $\mathfrak L$ together with $d^2=0$, the Leibniz rule, associativity and
commutativity in $\mathcal A$. Moreover, a strict morphism $\mathfrak L\to\mathfrak L'$ induces a strict morphism
$\mathcal A\otimes\mathfrak L\to\mathcal A\otimes\mathfrak L'$, and a differential ideal $J$ of $\mathcal A$ gives an $L_\infty$ ideal
$J\otimes\mathfrak L$. This follows because  every bracket with an argument in $J\otimes\mathfrak L$ has a factor in $J$.

Next, we will use
the above  statements for tensor products over $\mathbb C$, in the following sense.
The holomorphic single copies $L_{\rm RW}$ and $\mathfrak H$ are taken to consist of objects whose coefficients
depend holomorphically on $z$, and $\Kk$ of antiholomorphic forms whose coefficients depend on $\bar z$.
A priori, their tensor product over $\mathbb C$ then consists of functions of independent variables $z$ and
$\bar z$, i.e.\ of functions on the complexification of $X$, which has twice the complex dimension of $X$.
This is analogous to the standard double copy, where the fields depend on two sets of coordinates
$x$ and $\bar x$, one for each single copy. There, however, the brackets of the doubled theory satisfy
the $L_\infty$ relations only up to terms proportional to $\square-\bar\square$, so that the $L_\infty$ algebra does
not even exist on the doubled space. One has to impose the level-matching condition $\square=\bar\square$ of
string theory, which is solved by going to the undoubled space $x=\bar x$ \cite{BCDH22}. In the present case no
such condition is needed, and the $L_\infty$ algebra
is already defined on the complexification, with $z$ and $\bar z$ independent. The real slice
is needed, however, for 
the inner product, as the cycle over which the top form is integrated.
A function that is holomorphic in both sets of coordinates  is determined by its
values on the real slice where $\bar z$ is the complex conjugate of $z$  
and  the derivatives $\partial_i$ and $\bar\partial_{\bar\imath}$ become the usual
 derivatives \eqref{eq:coords}. Restricting to the real slice is therefore compatible with
all operations, and the tensor product describes real-analytic fields on $X$. General fields are
obtained by completion, for which all relations  continue to hold.

Let us then apply the above tensor product constructions to the single copies of the previous subsection. 
This gives the $L_\infty$ algebra $A=\Kk\otimes L_{\rm RW}$, with brackets $\ell_n^A$ as in \eqref{eq:tensorprinciple}, and the dg Lie
algebra $B=\Kk\otimes\mathfrak H$, with differential $d_B$ and brackets $[\,\cdot\,, \,\cdot\, ]_B$. 
Furthermore, $\Phi=1\otimes\Pi:A\to B$ defines a strict  $L_\infty$ morphism, and we have the ideals
$A^+=\Kk^+\otimes L_{\rm RW}$ and $B^+=\Kk^+\otimes\mathfrak H$. Concretely, $A$ consists of the forms $\Omega^{p,q}$ with
$p\le1$, in degree $p+q-1$, while $B$ consists of the forms $\Omega^{p,q}$ with $p\ge2$, in degree $p+q-2$.\footnote{Here and in the
following, $\Kk^q\otimes\Omega^{p,0}$ is identified with $\Omega^{p,q}$ by writing the holomorphic one-forms first.}
Every element of $A$ is a sum of components of definite antiholomorphic degree, and we split it
uniquely as $a=a^-+a^+$, where $a^-$ collects the components of antiholomorphic degree at most one and
$a^+$ those of degree at least two. Thus $a^-\in A^-:=\Kk^-\otimes L_{\rm RW}$ and $a^+\in A^+$, so that
$A=A^-\oplus A^+$ as a vector space, and $(\;\cdot \;)^\pm$ denote the projections onto the two summands. We
use the same notation for $B=B^-\oplus B^+$, with $B^-:=\Kk^-\otimes\mathfrak H$. Note that $A^+$ and $B^+$ are ideals,
while $A^-$ and $B^-$ are merely subspaces (a bracket of elements of $A^-$ can have components in $A^+$). 
Since $\Phi$ does not act on the  antiholomorphic factor, it maps $A^\pm$ into $B^\pm$.

The subspace $A^-$  consists of $\chi$, $\lambda$, $\bar\lambda$ and $e$, i.e.\ of the left half
of the folded complex, while the ideal $B^+$ consists of the forms $e^\ast$, $\lambda^\ast$, $\bar\lambda^\ast$ and $\chi^\ast$ in $\Omega^{2,2}$, $\Omega^{2,3}$, $\Omega^{3,2}$ and
$\Omega^{3,3}$, i.e.\ of the right half of the folded complex. The
double copy combines the two halves,
\be\label{eq:Fsplit}
 \mathcal F=A^-\oplus B^+=\big(\Kk^-\otimes L_{\rm RW}\big)\oplus\big(\Kk^+\otimes\mathfrak H\big)\;,
\ee
which is precisely the folded complex of section~\ref{sec:bicx}. We will next construct its $L_\infty$ brackets, 
proceeding  in two steps.

First, the ideal $B^+$ is acted on by $A$ through the embedding tensor: an element $a\in A$ acts on $b\in B^+$
by $[\Phi(a),b]_B$, where $[\cdot,\cdot]_B$ denotes the bracket of the dg Lie algebra $B$, i.e.\ the bracket of $\mathfrak H$
extended to forms as in \eqref{eq:tensorprinciple}. Since $B^+$ is an ideal of $B$, this lies again in $B^+$. Accordingly, the space $T=A\oplus B^+$ carries the brackets of a
semidirect product,
\be\label{eq:Tbrackets}
\begin{split}
 \mu_1(a+b)&=\ell_1^A(a)+d_B(b)\;,\\
 \mu_2(a_1+b_1,a_2+b_2)&=\ell_2^A(a_1,a_2)+[\Phi a_1,b_2]_B+[b_1,\Phi a_2]_B+[b_1,b_2]_B\;,\\
 \mu_3(a_1+b_1,a_2+b_2,a_3+b_3)&=\ell_3^A(a_1,a_2,a_3)\;,
\end{split}
\ee
for $a,a_i\in A$ and $b,b_i\in B^+$, and $\mu_{n\ge4}=0$. They satisfy the $L_\infty$ relations: the part in $A$
consists of the relations of $A$, while the part in $B^+$ follows from the Jacobi identity and the
derivation property of $d_B$ in $B$, because $\Phi$ commutes with the differentials, intertwines the binary
brackets, $\Phi(\ell_2^A(a_1,a_2))=[\Phi a_1,\Phi a_2]_B$, and annihilates $\ell_3^A$.

The above  space $T=A\oplus B^+$ is larger than $\mathcal F$, since it contains the full  $A$. In the second step we
therefore identify the `high part' of $A$ with its image under the embedding tensor by  dividing  by
the ideal 
\be
 I=\big\{a-\Phi(a)\ :\ a\in A^+\big\}\;\subset\;T\;.
\ee
This is indeed an ideal of $T$, which can be seen as follows. Note first that $I\subset T$, since
$\Phi(A^+)\subset B^+$. 
We have to show that every bracket \eqref{eq:Tbrackets} with at least one argument in $I$
lies again in $I$, i.e.\ is of the form $c-\Phi(c)$ with $c\in A^+$. Let $a\in A^+$ and let $a'+b'$, $a''+b''$ be
arbitrary elements of $T$. For the differential, using that $\Phi$ is a chain map, $d_B\Phi=\Phi\,\ell_1^A$, we compute 
\be\label{eq:idealmu1}
 \mu_1\big(a-\Phi(a)\big)=\ell_1^A(a)-d_B\Phi(a)=\ell_1^A(a)-\Phi\big(\ell_1^A(a)\big)\;,
\ee
which is of the required form with $c=\ell_1^A(a)$. For the binary bracket, \eqref{eq:Tbrackets} with
$a_1=a$, $b_1=-\Phi(a)$, $a_2=a'$ and $b_2=b'$ gives
\be\label{eq:idealmu2}
\begin{split}
 \mu_2\big(a-\Phi(a),a'+b'\big)&=\ell_2^A(a,a')+[\Phi a,b']_B+[-\Phi a,\Phi a']_B+[-\Phi a,b']_B\\
 &=\ell_2^A(a,a')-[\Phi a,\Phi a']_B\\
 &=\ell_2^A(a,a')-\Phi\big(\ell_2^A(a,a')\big)\;.
\end{split}
\ee
Here, in the second step, the action $[\Phi a,b']_B$ of $a$ on $b'$ has cancelled against the bracket
$[-\Phi a,b']_B$ of the $B^+$-components, and in the last step we used that $\Phi$ intertwines the binary
brackets. This is of the required form with $c=\ell_2^A(a,a')$. 
Finally, the ternary bracket only involves the components
in $A$,
\be\label{eq:idealmu3}
 \mu_3\big(a-\Phi(a),a'+b',a''+b''\big)=\ell_3^A(a,a',a'')=\ell_3^A(a,a',a'')-\Phi\big(\ell_3^A(a,a',a'')\big)\;,
\ee
since $\Phi$ annihilates the values of $\ell_3^A$; this  is of the required form with $c=\ell_3^A(a,a',a'')$. In all
three cases $c\in A^+$, because $A^+$ is an ideal of $A$.

Having established that $I$ is an ideal it follows that  $T/I$ is an $L_\infty$
algebra. In $T/I$ every element $a+b$ is equivalent to $a^-+\big(b+\Phi(a^+)\big)$, which has a unique
representative in $A^-\oplus B^+=\mathcal F$, so that $T/I\cong\mathcal F$. Computing the brackets \eqref{eq:Tbrackets} on
such representatives and passing to the representative of the result gives the brackets of $\mathcal F$:
for $a,a_i\in A^-$ and $b,b_i\in B^+$,
\be\label{eq:Fbrackets}
\begin{split}
 \ell_1(a+b)&=\big(\ell_1^A(a)\big)^-+\Phi\big((\ell_1^A(a))^+\big)+d_B(b)\;,\\
 \ell_2(a_1+b_1,a_2+b_2)&=\big(\ell_2^A(a_1,a_2)\big)^-+\Phi\big((\ell_2^A(a_1,a_2))^+\big)
 +[\Phi a_1,b_2]_B+[b_1,\Phi a_2]_B\;,\\
 \ell_3(a_1+b_1,a_2+b_2,a_3+b_3)&=\big(\ell_3^A(a_1,a_2,a_3)\big)^-\;,
\end{split}
\ee
and $\ell_{n\ge4}=0$. Here $[b_1,b_2]_B$ no longer appears, since the antiholomorphic degrees of two elements of
the right half add up to at least four.

These arguments prove the following proposition.

\begin{prop}\label{prop:Fdc}
$\mathcal F$ with the brackets \eqref{eq:Fbrackets} is an $L_\infty$ algebra. Its relations follow from the facts
that $\Kk$ is a graded commutative differential graded algebra with differential ideal $\Kk^+$, that $L_{\rm RW}$
is an $L_\infty$ algebra, that $\mathfrak H$ is a dg Lie algebra, and that $\Pi$ is a strict morphism.
\end{prop}

Let us spell out these brackets.
 The differential is $\partial$ and $\dbar$ within both halves, while the
bridge is obtained by first applying $\dbar$ and then the embedding tensor $\partial$, which reproduces
\eqref{eq:l1-folded}. The brackets of two elements of the left half that stay in the left half are the
brackets of $A$, i.e.\ those of the Roytenberg--Weinstein algebra with the antiholomorphic forms
multiplied,
\be\label{eq:Fleft}
\begin{split}
 &\ell_2(\lambda_1,\lambda_2)=[\lambda_1,\lambda_2]_C\,,\qquad
 \ell_2(\lambda,e)=[\lambda,e]_C\,,\qquad
 \ell_2(\lambda,\chi)=\tfrac12\,\iota_{\vth(\lambda)}\partial\chi\,,\\
 &\ell_2(\lambda,\bar\lambda)=\tfrac12\,\Lie_{\vth(\lambda)}\bar\lambda\,,\qquad
 \ell_2(\chi,e)=-\tfrac12\,\iota_{\vth(e)}\partial\chi\,.
\end{split}
\ee
The only bracket of two elements of the left half that is carried across is
\be\label{eq:Fhh}
 \ell_2(e_1,e_2)=\Phi\big(\ell_2^A(e_1,e_2)\big)=\partial[e_1,e_2]_C=[\partial e_1,\partial e_2]_B\;,
\ee
where the last step is the strictness of $\Pi$. The brackets of an element of the left half with one of
the right half are
\be\label{eq:Fmixed}
 \ell_2(\lambda,y^\ast)=[\partial\lambda,y^\ast]_B=\Lie_{\vth(\lambda)}y^\ast\;,\qquad
 \ell_2(e,e^\ast)=[\partial e,e^\ast]_B\;,\qquad
 \ell_2(e,\bar\lambda^\ast)=[\partial e,\bar\lambda^\ast]_B\;,
\ee
with the bracket $[\cdot,\cdot]_B$ of $B$. Finally, the ternary brackets
are those of $A$ that stay in the left half,
\be\label{eq:Fl3}
 \ell_3(\lambda_1,\lambda_2,\lambda_3)=-T(\lambda_1,\lambda_2,\lambda_3)\;,\qquad
 \ell_3(\lambda_1,\lambda_2,e)=-T(\lambda_1,\lambda_2,e)\;.
\ee
With \eqref{eq:Hbrackets} and $\partial\alpha=\iota_{\vth(\alpha)}\omega$, the brackets \eqref{eq:Fhh} and \eqref{eq:Fmixed} are those of
\eqref{eq:minl2}, the bracket of $\mathfrak H$ on $\Omega^{2,\bullet}$ being the Lie bracket of vector-valued forms transported
by $\omega$, i.e.\ minus the Schouten bracket. Hence $\mathcal F$ coincides with the $L_\infty$ algebra $L$ of
section~\ref{sec:bicx}, and Proposition \ref{prop:Fdc} proves Proposition \ref{prop:minimal}. Together with
Proposition \ref{prop:master} this also proves the master equation for \eqref{eq:master}.

\subsection{Including cohomology and moduli}\label{sec:cohom}

So far we have ignored global issues. The prepotential \eqref{eq:prepot} solves the constraint $\partial A'=0$
precisely if every $\partial$-closed $(2,1)$-form is $\partial$-exact. The obstruction to this is the  cohomology
$H^{2,1}$, which vanishes on $\mathbb C^3$ and some other manifolds of interest for  twisted holography \cite{CG}. 
On a compact Calabi--Yau threefold, Hodge theory implies that every $\partial$-closed $(2,1)$-form is the sum of a $\partial$-exact
form and a unique harmonic form $t$, where  the latter cannot be written as $\partial e$ \cite{Huybrechts}. 
We include these harmonic
$(2,1)$-forms, which describe the
infinitesimal deformations of the complex structure (complex structure moduli) 
\cite{Tian,Todorov,CdlO}, by enlarging the folded complex by the harmonic
$(2,1)$-forms and their duals,
\be\label{eq:Lext}
 L^{\rm ext}=L\;\oplus\;\mathcal H^{2,1}\;\oplus\;\mathcal H^{1,2}\;,
\ee
with $t\in\mathcal H^{2,1}$ in degree $1$ and $t^\ast\in\mathcal H^{1,2}$ in degree $2$, extending the differential by zero and the
inner product by $\cip{t}{t^\ast}=\int_Xt^\ast\wedge t$. The full Beltrami differential is simply shifted as 
\be\label{eq:Afull}
 A=\vth(e)+\hat t\;, 
\ee
where we note that also $\hat t$ is divergence-free, since harmonic forms on a compact
K\"ahler manifold are $d$-closed, so that $\partial t=0$, and \eqref{eq:divomega} gives $(\dv\hat t)\lrcorner\omega=-\partial t=0$.
Equivalently, $A'=\partial e+t$ satisfies the constraint $\partial A'=0$.\footnote{On a compact Calabi--Yau threefold, the prepotential
$e$ also has harmonic $(1,1)$ zero modes, so that shifts
$e\rightarrow e+k$, with $k\in\mathcal H^{1,1}$, leave
$A'=\partial e+t$ and the classical action unchanged,
but are not generally generated by the linearized gauge
transformations. We implicitly fix these additional zero
modes and leave a complete treatment of global BV zero
modes aside.}

The brackets $\ell_2(e,e)$, $\ell_2(e,e^\ast)$ and $\ell_2(e,\bar\lambda^\ast)$ of $L$ depend on $e$ only through $\vth(e)$, and the
substitution $\vth(e)\to\vth(e)+\hat t$ gives the new brackets
\be\label{eq:minext}
\begin{split}
 \ell_2(e,t)&=-\{\vth(e),\hat t\}\lrcorner\omega\;,\qquad
 \ell_2(t_1,t_2)=-\{\hat t_1,\hat t_2\}\lrcorner\omega\;,\\
 \ell_2(t,e^\ast)&=\{\hat t,e^\ast\lrcorner\omega^{-1}\}\lrcorner\omega\;,\qquad
 \ell_2(t,\bar\lambda^\ast)=\partial\big((\hat t\wedge c)\lrcorner\omega\big)\;.
\end{split}
\ee
The remaining brackets of $L$ involving $e$ would require a prepotential for $t$, which does not exist
globally. They are replaced by
\be\label{eq:xit}
 \ell_2(\lambda,t)=\iota_{\vth(\lambda)}t-\partial\,\iota_{\hat t}\lambda\;,\qquad
 \ell_2(\chi,t)=-\iota_{\hat t}\partial\chi\;,
\ee
while no ternary bracket involving $t$ is needed, and all brackets with an argument $t^\ast$ vanish.

\begin{prop}\label{prop:minext}
The brackets of $L$ together with \eqref{eq:minext} and \eqref{eq:xit} satisfy all $L_\infty$ relations with values
in $L$.
\end{prop}

\noindent
This follows from the double copy as before, with the left half enlarged by $\mathcal H^{2,1}$: the harmonic
forms lie in $B^-$, but not in the image of $\Phi$, and one extends $\Phi$ by $\Phi(t)=t$. In particular,
$\partial\,\ell_2(\lambda,t)=\partial\,\iota_{\vth(\lambda)}t=\Lie_{\vth(\lambda)}t=[\partial\lambda,t]$ by Cartan's formula and $\partial t=0$, while $\ell_2(\chi,t)$ lies in
the kernel of $\Phi$. We have also verified the relations explicitly, using that locally every harmonic
$(2,1)$-form is of the form $\partial\dbar\beta$. The brackets with values in $\mathcal H^{1,2}$ are fixed by cyclicity.

The field equation becomes $\partial\dbar e-\tfrac12\{A,A\}\lrcorner\omega=0$,
i.e.\ the Kodaira--Spencer equation for the full Beltrami differential
\eqref{eq:Afull}, and the gauge transformations become
$\delta e=\dbar\lambda+\partial\bar\lambda-[\lambda,e]_C
-\iota_{\vth(\lambda)}t+\partial\,\iota_{\hat t}\lambda$.
The isomorphism of Proposition \ref{prop:master} extends to the moduli
without any further components. Although the original $L_\infty$ algebra
has no ternary brackets involving $t$, the isomorphism generates the
ternary bracket
$\ell_3^{\rm cyc}(\lambda_1,\lambda_2,t)=Q(\lambda_1,\lambda_2,t)$,
where
\be 
\begin{aligned}
Q(\lambda_1,\lambda_2,t)
={}&\iota_{\vth(\lambda_1)}\iota_{\vth(\lambda_2)}t
-\Lie_{\vth(\lambda_1)}\iota_{\hat t}\lambda_2
+\Lie_{\vth(\lambda_2)}\iota_{\hat t}\lambda_1
+\iota_{\hat t}[\lambda_1,\lambda_2]_C\,.
\end{aligned}
\ee 
The cyclic algebra is then encoded in the master action
\be\label{eq:masterext}
\begin{aligned}
 S_{\rm ext}=\int_X\Big[\;
 &\tfrac12\,e\wedge\partial\dbar e
 -e^\ast\wedge\big(\dbar\lambda+\partial\bar\lambda-\Lie_{\vth(\lambda)}e-\iota_{\vth(\lambda)}t+\partial\,\iota_{\hat t}\lambda\big)
 +\lambda^\ast\wedge\big(\partial\chi+\tfrac12[\lambda,\lambda]_C\big)\\
 &+\chi^\ast\wedge\big(\tfrac12\,\iota_{\vth(\lambda)}\partial\chi+\tfrac16\,T(\lambda,\lambda,\lambda)\big)
 +\bar\lambda^\ast\wedge\big(\dbar\chi-\iota_{\hat t}\partial\chi+\Lie_{\vth(\lambda)}\bar\lambda\\
 &\qquad\qquad+\tfrac12\,\ell_2(\lambda,\lambda)\big|_{\Omega^{0,1}}+\tfrac12\,Q(\lambda,\lambda,t)\big)\Big]
 +\tfrac16\int_X\omega\wedge\big((A\wedge A\wedge A)\lrcorner\omega\big)\;,
\end{aligned}
\ee
with $A$ as in \eqref{eq:Afull}. For $t=0$ this reduces to
\eqref{eq:master}. 
The cubic term now contains the Yukawa coupling
$\tfrac16 C_{abc}t^at^bt^c$, with
$C_{abc}=\int_X\omega\wedge
((\hat t_a\wedge\hat t_b\wedge\hat t_c)\lrcorner\omega)$.

The moduli enter \eqref{eq:masterext} without kinetic term and without gauge transformation, and $t^\ast$
does not appear, so that $S_{\rm ext}$ is best viewed as a family of BV theories for the local fields,
parametrized by $t$. Varying $t$ is not part of Kodaira--Spencer theory: by the Tian--Todorov lemma
\cite{Tian,Todorov}, which in our conventions follows directly from \eqref{eq:koszul} as
$\{\beta,\gamma\}=(-1)^\beta\dv(\beta\wedge\gamma)$ for $\dv\beta=\dv\gamma=0$, the bracket $\{A,A\}$ has no harmonic part, so that the
harmonic part of the Kodaira--Spencer equation vanishes identically and the moduli are unobstructed.
They are thus background data, as in \cite{BCOV}.

\section{Conclusions and outlook}\label{sec:conclusions}

In this paper we have presented a local action principle for Kodaira--Spencer gravity. 
To this end  the constraint $\partial A'=0$, on which the non-local action of BCOV is based \cite{BCOV93,BCOV}, 
is solved  in terms of a $(1,1)$-form $e$ by writing  $A' = \partial e$. 
This construction is a natural generalization of  \cite{HS}, which gave a 
local reformulation of Bandos--Townsend theories \cite{Bandos:2008jv}. 
Remarkably, 
apart from making the action local, 
this induces at the level of the free theory precisely the gauge transformations of double field theory. 
Moreover, the non-linear gauge structure is based on a Courant algebroid and its associated 
$L_{\infty}$ algebra, precisely as in standard double field theory. 
In addition, the $L_{\infty}$ algebra defining the full theory has a  double copy interpretation
that is analogous to \cite{Bonezzi:2022bse,BCDH22}. 
The present theory differs substantially 
from  standard double field theory, however, in that its classical action only has cubic couplings, and no level-matching 
or section constraint is needed  for consistency.

The present formulation clarifies in which sense Kodaira--Spencer theory is really a theory 
of gravity (albeit a topological one). Concretely, we truncated the theory to the three-dimensional 
real slice $z=\bar{z}$ for which the gauge transformations reduce to the spin-2 and B-field gauge transformations, while 
the quadratic action reduces to the sum of Fierz--Pauli action for massless spin-2 and the 
action for a Kalb--Ramond B-field. At the non-linear level the theory differs, however, from ordinary 
double field theory or Einstein--Hilbert gravity coupled to a B-field. Indeed, the theory is exact with 
only cubic couplings, which carry three derivatives and violate parity, in contrast to the parity even 
and non-polynomial two-derivative interactions of Einstein--Hilbert theory. 
To cubic order, and truncating the antisymmetric B-field, the action coincides with 
an exotic 3D gravity coupling found by Boulanger and Gualtieri \cite{BG}. Integrating out the B-field 
leads to non-localities that, however, can be removed by non-local field redefinitions, at least up to and including  quartic order.

Kodaira--Spencer gravity has also been related to three-dimensional Einstein gravity in recent work. In particular, Erdmenger et al.\ studied Kodaira--Spencer gravity on $SL(2,\mathbb C)\cong H^3\times S^3$ \cite{EKKMS25,EKKS26}, a background relevant for twisted holography \cite{CG}. They showed that, upon restricting to a subsector of complex structure deformations defined by a `locality constraint' stronger than $\partial A'=0$, the prepotential can be expressed \textit{algebraically} in terms of the Beltrami differential, up to $\partial$-closed terms. This leads to a holomorphic Chern--Simons action for $SL(2,\mathbb C)$, whose dimensional reduction yields chiral $AdS_3$ gravity. It would be interesting to investigate this relation further using the present formulation. (The cohomological obstruction discussed in section~\ref{sec:cohom} vanishes on this background \cite{Hoermander}.)

Furthermore, Ben-Shahar et al.\ analyzed a double copy of 3D Chern--Simons theory
and showed that it is on-shell equivalent to 3D gravity in a unimodular gauge \cite{BBZ26},
see also \cite{BBZ25}, with the
volume-preserving diffeomorphisms as gauge symmetries. In terms of a
divergence-free field $A^\mu{}_a$, the action has a non-local quadratic term and a
cubic term, while a local formulation in terms of constrained vector frames is also available.
This is closely related to the real version of the non-local action \eqref{eq:nonlocal}:
identifying the holomorphic vector index of the Beltrami differential with $\mu$ and its antiholomorphic
form index with $a$, and correspondingly $\bar z$ with $z$, one recovers the double copy action of
\cite{BSJ}, generalized in \cite{BDH}. Our prepotential then provides a local
parametrization of the divergence-free field, as seen in the reduction of
section~\ref{sec:3d}.

The present work should be extended in various directions. Most intriguingly, it may provide a powerful 
toy model to study how gravity may be completed to a theory in higher dimensions with an improved UV behavior.  
Indeed, the 3D gravity theory found here through dimensional reduction is badly non-renormalizable by power counting (although it has no local degrees of freedom), yet its 
higher-dimensional 6D origin in Kodaira--Spencer gravity may have a well-defined quantization. 
Indeed, Costello and Li developed the perturbative quantization of BCOV theory in the BV formalism \cite{CL,CL2}. 
It would be interesting to study such features in the present formulation, which is more `gravity-like', and to see whether 
it may shed  light on possible quantizations of non-topological gravity. Finally, it would be interesting to study 
twisted holography in the present formulation.

\section*{Acknowledgments and use of AI}

The main idea of this paper is mine, but everything else is 
heavily based on interactions with artificial intelligence, particularly with Claude Opus 5.5, 
which I used in all aspects: for the brainstorming of ideas, literature search, explicit computations, and the write-up 
of results. For instance, the double copy interpretation of section~4 is essentially due to Claude, although the final presentation 
went through countless iterations, transforming  an initially highly involved   mathematical proposal to a reasonably 
simple one. Moreover, any AI-written parts of the text went through many iterations to hopefully eliminate  all instances 
of `AI slop'.  
After completing a first final draft I used ChatGPT (GPT-6) for a critical review of the entire manuscript and in particular 
 to independently derive and check the $L_\infty$  brackets transported by the isomorphism \eqref{eq:iso}, including the absence of higher brackets, and to verify the cyclicity condition \eqref{eq:cycsign}. 
I take  responsibility for the entire paper.

\medskip

Moreover, I would like to thank my human colleagues  Roberto Bonezzi, Nicolas Boulanger, Christoph Chiaffrino and Barton Zwiebach for discussions and comments on this project, and Kevin Costello for alerting me to ref.~\cite{Saberi:2020pmw}.

\end{document}